\documentclass[reprint,amsmath,amssymb,aps,prc,nofootinbib,superscriptaddress,floatfix]{revtex4-1}
\usepackage{bm}
\usepackage{graphicx}
\usepackage{dcolumn}
\usepackage{color}
\newcommand{\be}{\begin{equation}}
\newcommand{\ee}{\end{equation}}
\newcommand{\bea}{\begin{eqnarray}}
\newcommand{\eea}{\end{eqnarray}}

\newcommand{\bfbj}{\mbox{\boldmath $J$}}

\newcommand{\mbss}[1]{_{\mbox{\scriptsize #1}}}

\newcommand{\mbsu}[1]{\mbox{\scriptsize #1}}

\newcommand{\vphu}{\vphantom{*}}

\newcommand{\hfm}{\hphantom{-}}
\definecolor{dgreen}{rgb}{0.0,0.5,0.0}

\begin{document}

\title{Skyrme RPA for nuclear resonances: trouble with magnetic modes}

\author{J. Speth}
\affiliation{Institut f\"ur Kernphysik, Forschungszentrum J\"ulich, D-52425 J\"ulich, Germany}
\author{P.-G. Reinhard}
\affiliation{Institut f\"ur Theoretische Physik II, Universit\"at Erlangen-N\"urnberg,
D-91058 Erlangen, Germany}
\author{V. Tselyaev}
\affiliation{St. Petersburg State University, St. Petersburg, 199034, Russia}
\email{tselyaev@mail.ru}
\author{N. Lyutorovich}
\affiliation{St. Petersburg State University, St. Petersburg, 199034, Russia}
\date{\today}

\begin{abstract}
We discuss major differences between electric and magnetic excitations
in nuclei appearing in self-consistent calculation based on Skyrme
energy-density functionals.  Tools of analysis are Landau-Migdal
parameters for bulk properties and RPA for resonance modes of
$^{208}$Pb as representative of finite nuclei.  We show that the
relation between the effective mass and the effective particle-hole
interaction, well known in the Landau-Migdal theory, explains the
success of self-consistent calculations of electric transitions in
such approaches. This effect, however, does not automatically exist in
the magnetic case. This calls for further developments of the
Skyrme functional in the spin channel.
\end{abstract}


\maketitle

\section{Introduction}
\label{sec:Intr}

Two different approaches have been successfully applied to calculation
of nuclear resonance excitations.  The traditionally most often used
method of the first kind starts with a phenomenological
single-particle model and an effective residual interaction. A widely
used and powerful version is Migdal's theory of \emph{Finite Fermi Systems}
\cite{Migdal_1967} based on Landau's
\emph{Theory of Fermi Liquids} \cite{LanLif9}. This approach has been
applied extensively to a broad range of nuclei, for reviews see
\cite{Speth_1977}.  In the second approach one starts with an
effective energy density functional (EDF) which allows to derive the
single particle model as well as the residual interaction. One of the
most often used version is the Skyrme-Hartree-Fock (SHF) approach
\cite{Sky59a,Neg72a,Neg75a,Liu76}, for a review on SHF see
\cite{Bender_2003}.  Originally it was designed as a model for the
nuclear ground state \cite{Brink72}. But soon it was also applied to
compute self-consistently collective excitation states, especially
giant resonances. The parameters of the early Skyrme EDF were
predominately adjusted to ground state properties which does not a
priory guarantee an appropriate particle-hole ($ph$)
interaction. However, as incompressibility and symmetry energy are
closely connected to the spin-independent isoscalar and isovector parts
of the $ph$-interaction, in general the theoretical results were not
so bad. In later parametrizations, also properties of excited states
were included which improved the theoretical results compared to the
data, see e.g. \cite{Kluepfel_2009}. The Skyrme EDF turned out to be
flexible enough to reproduce all collective modes with natural parity.
Most of the modern parameter sets use very similar values for the
incompressibility $K_{\infty}$ and symmetry energy $a_\mathrm{sym}$.
There is
more variation in the choice of the effective mass $m^*$. As the
effective mass has a heavy impact on the $ph$-spectrum, it is somewhat
surprising that the RPA results of giant resonances and also of
low-lying collective states are not very different.  In that respect,
it is instructive to learn from Landau Migdal (LM) theory; there it is
known that the (spin-independent) isoscalar and isovector force
parameters $f_0$ and $f'_0$ are correlated with the $f_1$ parameter
quantifying effective mass, a feature similar to the \emph{backflow}
of quasi-particles in condensed matter \cite{Bardin_1961}.
With decreasing effective masses the isoscalar
$ph$-interaction gets more attractive and the isovector $ph$-interaction
becomes less repulsive thus correcting for the larger $ph$-energy
spacing. We will show that this effect exists also for self-consistent
calculations in the framework of Skyrme EDF.

Unfortunately there are no spin magnetic \emph{bulk properties} known
which are directly related to the spin dependent Landau-Migdal
parameter $g_0$ and $g'_0$.  Moreover, there exists only one truly
collective spin mode, namely the Gamow-Teller (GT)-resonance in
neutron rich nuclei, which is related to the spin-isospin part of the
residual $ph$-interaction.  Therefore, the parameters which are most
relevant for the spin-dependent part of the $ph$-interaction of the
existing parametrizations were not yet adjusted to experimental
properties with the exception of the two-body spin-orbit
interaction. Bell and Skyrme introduced already more then 50 years ago
a two-body spin-orbit term into the original ansatz in order to
reproduce the single particle (sp) spin-orbit splitting
\cite{Bell_1956}. Van Giai and Sagawa \cite{vanGiai_1981} modified two
Skyrme parametrizations where they considered the spin-dependent
LM-parameter $g_0$ and $g'_0$ as additional constraints and calculated
GT states in some doubly magic nuclei. Self-consistent calculations
where the spin-dependent $ph$-interaction plays a role e.g. magnetic
excitations, are very scarce. They came up only recently
\cite{Vesely_2009,Nesterenko_2010,Tselyaev_2019} and point toward
insufficiency's of the Skyrme EDF as given.  In the survey
\cite{Tselyaev_2019}, the spin-relevant parameters of the Skyrme EDF
were modified to reproduce the experimental data which amounts to a
substantial readjustment of the LM parameters $g_0$, $g'_0$.

The aim of this paper is to analyze the success of conventional Skyrme
EDF for the natural parity (also called electrical) collective
resonances and the failure in the spin channel.  We use as tool the
trends of LM parameters with effective mass. We start with the well
settled and well working case of the electrical modes where we find
that the``backflow effect'' on the $ph$-interaction is the key to
success. Then we apply the same strategy to LM parameters in the spin
channel.  In section \ref{sec:Skyrme} we give a short review into the
Skyrme approach. In section \ref{sec:LMT} we present the relevant
formulas of the Landau-Migdal approach where we especially emphasize
the connection of the LM parameter with nuclear matter properties. In
this connection we analyze the functional behavior of the
spin-dependent and spin-independent LM parameter on the effective
mass. Section \ref{sec:electric} contain the main result of our
investigation. First we compare RPA and TBA results of electrical
giant resonances with unperturbed $ph$ energies for various Skyrme
parametrizations. In the second part we show the corresponding results
for the magnetic states using as example the $1^+$ in
$^{208}$Pb. Finally we summarize and discuss our results in section
\ref{sec:sum}. In the Appendix \ref{sec:densdepLM} we discuss the
density-dependence of $f_0$ and $f'_0$ which is crucial in the LM
theory. As supplement to the section on giant resonances, we present
in Appendix \ref{sec:low} RPA results for low-lying collective states.
Appendices \ref{app:nmp} and \ref{app:lmp} contain the known
formulas expressing the nuclear matter properties and the LM parameters
in terms of the parameters of the Skyrme EDF.

\section{The Skyrme energy functional}
\label{sec:Skyrme}

The Skyrme energy functional consists of kinetic energy, Coulomb
energy, pairing energy, and, as key entry, the Skyrme interaction
energy. This is well documented in several reviews, see
e.g. \cite{Bender_2003,Erler_2011}. We recall here just the core piece as
far as  is needed in following.  The Skyrme interaction energy is
formulated in terms of a few nuclear densities and currents as are:
density $\rho_T$, kinetic density $\tau_T$, spin-orbit density
$\vec{J}_T$, spin-tensor density $\mathbb{J}_T$, current $\vec{j}_T$,
spin density $\vec{s}_T$, and spin kinetic density, where the
index $T$ stands for isospin ($T=0$ or 1). It reads in commonly used
form
\begin{subequations}
\label{eq:basfunct}
\begin{eqnarray}
{\cal E}_{\rm Sk}
& = & \sum_{T = 0,1} 
      \Big(   {\cal E}_{T}^{\rm even}
            + {\cal E}_{T}^{\rm odd}
      \Big)
      \;,
\\
  {\cal E}_{T}^{\rm even}
  & = &
        C_T^{\rho}(\rho_0) \, \rho_{T}^{2}
      + C_T^{\Delta \rho} \, \rho_{T} \Delta \rho_{T}
      + C_T^{\tau} \, \rho_{T} \tau_{T}
\nonumber\\
  &&
      + C_T^{\nabla J} \rho_{T} \, \nabla\!\cdot\!\vec{J}_{T}
      \;(+ C_T^{J} \mathbb{J}^2_{T} )
     \;,
\label{eq:Eskeven}
\\
  {\cal E}_{T}^{\rm odd}
  & = &
  C_T^{s}(\rho_0) \, \vec{s}^{2}_{T}
      + C_T^{\Delta s} \, \vec{s}_{T}\!\cdot\!\Delta \vec{s}_{T}
 \nonumber \\
  &   &
      + C_T^{j} \, \vec{j}^2_{T}
       + C_T^{\nabla j} \,
        \vec{s}_{T} \!\cdot\! \nabla\!\times\! \vec{j}_{T}
      \;(+ C_T^{sT} \, \vec{s}_{T} \!\cdot\! \vec{\tau}_{T} )
     .
\label{eq:Eskodd}
\end{eqnarray}
\end{subequations}
The terms employing the tensor spin-orbit densities are written in
brackets to indicate that these terms are ignored in the majority of
published Skyrme parametrizations.  Only the time even part
$\mathcal{E}_{T}^{\rm even}$ is relevant for ground states of
even-even nuclei. Time-odd nuclei and magnetic excitations are
sensitive also to the time-odd part $\mathcal{E}_{T}^{\rm odd}$.  The
parameters $C_T^\mathrm{type}$ for each term in the time-even part are
adjusted independently, usually to a carefully chosen set of empirical
data \cite{Bender_2003,Erler_2011}.  A couple of different options are
conceivable for the parameters of the time-odd terms which has
consequences for the description of magnetic modes. This will be
discussed in section \ref{sec:mag}.

The original formulation of the Skyrme-Hartree-Fock (SHF) method was
based on the concept of an effective interaction, coined the Skyrme
force \cite{Sky59a}. Modern treatments of SHF, however, start from a
Skyrme energy-density functional as shown above.  Nonetheless, the
Skyrme force was the original motivation to develop the Skyrme
functional and, being a zero-range interaction, displays an obvious
similarity to the Landau-Migdal force.  Its interaction part without
spin-orbit and Coulomb terms has the form
\begin{subequations}
\label{eq:skenfun}
\begin{eqnarray}
  E^{\mbsu{int}}_\mathrm{Sk}
  &=&
  E^{\mbsu{int}}_\mathrm{Sk,dens}
  +
  E^{\mbsu{int}}_\mathrm{Sk,grad}\,,
\\
 E^{\mbsu{int}}_\mathrm{Sk,dens}
  &=&
  \langle\Phi|
  t_0(1\!+\!x_0 \hat{ P}_\sigma)\delta(\mathbf{r}_{12})
\nonumber\\
  &+& \frac{t_3}{6}(1\!+\!x_3\hat P_\sigma)\rho^\alpha\left(\mathbf{r}_1\right)
  \delta(\mathbf{r}_{12})|\Phi\rangle,
\label{eq:skenfun1}\\
 E^{\mbsu{int}}_\mathrm{Sk,grad}
  &=&
    \langle\Phi|
   \frac{t_1}{2}(1\!+\!x_1\hat{P}_\sigma)
  \left(
   \delta(\mathbf{r}_{12})\hat{\boldsymbol k}^2
   +
   {\hat{\boldsymbol{k}}}^{2}\delta(\mathbf{r}_{12})
  \right)
\nonumber\\
  &+& t_2(1\!+\!x_2\hat P_\sigma)\hat{\boldsymbol k}
  \delta(\mathbf{r}_{12})\hat{\boldsymbol k}
  |\Phi\rangle,
\label{eq:skenfun2}
\end{eqnarray}
\end{subequations}
where $\mathbf{r}_{12}=\mathbf{r}_1-\mathbf{r}_2$ and $\hat{P}_\sigma=
\frac{1}{2}(1+\hat{\boldsymbol \sigma}_1\hat{\boldsymbol{\sigma}}_2)$
is the spin-exchange operator. The $\hat{\boldsymbol k}$ stand for the
momentum operators.

The Skyrme interaction (\ref{eq:skenfun}) is not to be mixed with the
residual interaction for computing excitation properties within RPA,
called henceforth $ph$-interaction. This residual interaction is
deduced as second functional derivative of the Skyrme energy functional
(\ref{eq:basfunct}) \cite{Rei92b} with respect to the local densities
and currents it contains.  As the functional (\ref{eq:basfunct}) is
composed of contact terms, the RPA residual interaction is a
zero-range interaction. In that respect, it is very similar to the
Landau-Migdal interaction, a feature which motivates a discussion of
Skyrme RPA excitations in terms of Landau-Migdal (LM) parameters as
we do here.

The Skyrme functional contains kinetic terms which leads to an
effective nucleon mass $m^*$ which differs from the bare mass $m$ in
the nuclear interior. This has consequences for many time-odd
observables.  For example, the current operator $\hat{\vec{j}}_{q}$
fails to satisfy the continuity equations if $m^*\neq m$. The
non-trivial kinetic terms in the mean-field Hamiltonian call for a
dynamical correction which reads
\cite{Rep19a}
\begin{eqnarray}
  \hat{\vec{j}}_{\mathrm{eff},q}
  &=&
  \hat{\vec{j}}_{q}
  +
  \frac{m_q}{\hbar^2} \Bigl(2b_1 \left[\rho_{\bar{q}}\hat{\vec{j}}_{q}
         -\rho_{q}\hat{\vec{j}}_{\bar{q}}\right]
\nonumber\\
  &+&
  b_4\left[
       \rho_{\bar{q}}\vec{\nabla}\times\hat{\vec{\sigma}}_{q}
       -
       \rho_{q}\vec{\nabla}\times\hat{\vec{\sigma}}_{\bar{q}}
        \right] \Bigr)
\label{eq:backflow}
\end{eqnarray}
where $q$ denotes proton or neutron, $\bar{q}$ the nucleon with
opposite isospin, the coefficients $b_1$ and $b_4$ are defined in
Ref. \cite{Rei92b}.
This correction is crucial, e.g., in the computation
of transition strengths for giant resonances \cite{Ring80}. It
exemplifies the backflow effect known from the theory of Fermi
liquids \cite{Pin66}. The same correction is also required
for the magnetic current \cite{McN86a}. We will see below that a
similar backflow-like correction appears also for the residual
interaction in RPA.

\section{Landau-Migdal Theory}
\label{sec:LMT}

The Landau-Migdal theory of excitations in fermionic systems was
developed originally in the context of Fermi fluids
\cite{Landau1,Landau2,Landau3} and extended later to finite nuclei
\cite{Migdal_1967}.  The LM $ph$-interaction is restricted to the
Fermi surface where it depends only on the angle between the momenta
$\mathbf{p}$ and $\mathbf{p'}$ of the $1ph$ states before and after
the collision.  The $ph$-interaction in momentum space is a function
$F^{ph}(\mathbf{p},\mathbf{p'})$ times spin and isospin operators. The
momentum dependence can be expanded in terms of Legendre polynomials
in the dimensionless combination
$\mathbf{p}\cdot\mathbf{p'}/|\mathbf{p}||\mathbf{p'}|$
\cite{Landau1,Landau2,Landau3}.  The coefficients of this expansion
are called LM parameters. The leading order ($l=0$) of the
  Legendre polynomial is a constant which gives rise in
  $\mathbf{r}$-space to a delta function
  ($F^{ph}(1,2)=\delta(\mathbf{r}_{12})$) similar as the leading term
  in the Skyrme interaction. The term next to leading order ($l=1$) is
  proportional to $(\mathbf{p}\!\cdot\!\mathbf{p'})$. To deal better
with the finite size of the nuclei, one often introduces density
dependent LM parameters in the following way \cite{Migdal_1967}:
\begin{equation*}
  f(\rho)
  =
  f^\mathrm{ex} + (f^\mathrm{in}-f^\mathrm{ex})\frac{\rho_0(r)}{\rho_0(0)}
\end{equation*}
where $f^\mathrm{ex}$ stands for the exterior region of the nucleus
and $f^\mathrm{in}$ for the interior. However, the density dependence
of the Skyrme $ph$-interaction differs from that form which would
require a discussion of its own \cite{Speth_2014}. Thus we concentrate
on the interior region, the nuclear bulk properties, and drop the
upper index ``in'' in the following.

\subsection{Dimensionless Landau-Migdal (LM) parameters}

The expansion parameters have the
same dimension as the interaction, namely energy$\times$length$^3$, which
varies strongly with system size. To obtain a dimensionless measure of
interaction strength, it is customary to single out a pre-factor
having this dimension. A natural measure of length$^3$ is the inverse
of bulk density $\rho_0$. Thus Migdal uses as normalization factor the
derivative
$C_0^\mathrm{(Migdal)}=d\epsilon_\mathrm{F}/d\rho_0=\pi^2\hbar^2/(mk_\mathrm{F})
\approx{300}\,\mathrm{MeV\,fm}^3$ which applies to models using bare
nucleon mass $m$ \cite{Migdal_1967}.  Landau et al. take a similar normalization factor,
however, half of that and keeping the effective
nucleon mass $m^*$ in the definition \cite{LanLif9}. This amounts to
parametrize the RPA interaction in terms of LM parameters $F_l$, $G_l$
as
\begin{subequations}
\label{eq:consistent}
\begin{eqnarray}
  F^{ph}(\mathbf{p},\mathbf{p'})
  &=&
  C_0^*\sum_{l} P_l(\frac{\mathbf{p}\!\cdot\!\mathbf{p'}}{ k^2_{\rm F} })
  \Bigl[F_l + F^{\prime}_l\mathbf{\tau_1} \!\cdot\!\mathbf{\tau_2}
\nonumber\\
  &&
  \quad
  +G_l\mathbf{\sigma_1} \!\cdot\!\mathbf{\sigma_2}
  +G^{\prime}_l
  \mathbf{\sigma_1}\!\cdot\!\mathbf{\sigma_2}\mathbf{\tau_1}
  \!\cdot\!\mathbf{\tau_2}\Bigr]
\label{eq:F-SHF}
\\
  C_0^*
  &=&
 \frac{\pi^2 \hbar^2}{2 m^* k_{\rm F} }
 \approx
 {150}\,\mathrm{MeV\,fm}^3\,\frac{m}{m^*}
 \;.
\label{eq:C0star}
\end{eqnarray}
\end{subequations}
This normalization has the advantage that the condition for stable RPA modes
becomes simply $F_0>-1$ and it is most suited for self-consistent
nuclear models where effective mass $m^*\neq m$ plays a role.  Many
publications in the context of the Skyrme-Hartree-Fock (SHF) model use
this normalization. Still, Migdal's definition using a fixed normalization factor
is also often used, particularly in the empirical LM model. Thus we
discuss both definition side by side.  However, we want to avoid the
trivial, but distracting, factor two in the comparison and use
for that the normalization form
\begin{subequations}
\label{eq:fixed}
\begin{eqnarray}
  F^{ph}(\mathbf{p},\mathbf{p'})
  &=&
  C_0\sum_{l} P_l(\frac{\mathbf{p}\!\cdot\!\mathbf{p'}}{ k^2_{\rm F} })
  \Bigl[f_l + f^{\prime}_l\mathbf{\tau_1} \!\cdot\!\mathbf{\tau_2}
\nonumber\\
  &&
  \quad
  +g_l\mathbf{\sigma_1} \!\cdot\!\mathbf{\sigma_2}
  +g^{\prime}_l
  \mathbf{\sigma_1}\!\cdot\!\mathbf{\sigma_2}\mathbf{\tau_1}
  \!\cdot\!\mathbf{\tau_2}\Bigr]
\label{eq:F-LM}
\\
  C_0
  &=&
 \frac{\pi^2 \hbar^2}{2 m k_{\rm F} }
 \approx
 {150}\,\mathrm{MeV\,fm}^3
 \;.
\label{eq:C0}
\end{eqnarray}
\end{subequations}
 Henceforth we call the choice (\ref{eq:fixed}) "bare-mass normalization" and the
 choice (\ref{eq:consistent}) "effective-mass normalization". Each one of
the two definitions has its advantages and disadvantages. The bare-mass
normalization (\ref{eq:fixed}) produces a measure of strength of residual
interaction term by term comparable across SHF parametrizations with
different $m^*/m$. The effective-mass normalization (\ref{eq:consistent})
produces comparable effects of the residual interaction (stability
condition, excitation energies). The reason is that different
interaction strengths are required to compensate the impact of
different $m^*/m$, similar as in the backflow effect \cite{Bardin_1961}, see
eq. (\ref{eq:backflow}).

\subsection{Relation to nuclear matter properties (NMP)}

\begin{table}
\begin{center}
\begin{tabular}{ll}
\hline
\multicolumn{2}{c}{LM parameters}
\\
\multicolumn{1}{c}{effective-mass normalization} &
\multicolumn{1}{c}{bare-mass normalization}
\\
\hline
  $\displaystyle F_0=\frac{K_{\infty}}{6T_\mathrm{F}^*} - 1$
  &
  $\displaystyle f_0=\frac{m}{m^*}F_0
   =\frac{K_{\infty}}{6T_\mathrm{F}} - \frac{m}{m^*}$
\\[6pt]
  $\displaystyle F'_0=\frac{3a_{\mathrm{sym}}}{T_\mathrm{F}^*} - 1$
  &
  $\displaystyle f'_0=\frac{m}{m^*}F'_0=
   \frac{3a_{\mathrm{sym}}}{T_\mathrm{F}} - \frac{m}{m^*}$
\\[6pt]
  $\displaystyle F_1=3\left(\frac{m^*}{m} - 1 \right)$
  &
  $\displaystyle f_1=\frac{m}{m^*}F_1=3\left(1-\frac{m}{m^*}\right)$
\\[6pt]
  \multicolumn{2}{l}{
  $\displaystyle
    F'_1=3\left((1\!+\!\kappa_\mathrm{TRK})\frac{m^*}{m}-1\right)\;$}
\\[6pt]
  \multicolumn{2}{r}{
  $\displaystyle\; f'_1=\frac{m}{m^*}F'_1=
    3\left(1\!+\!\kappa_\mathrm{TRK}-\frac{m}{m^*}\right)\;$
  }
\\
\hline
\end{tabular}
\\[6pt]
\begin{tabular}{lcc}
\hline
 NMP & \multicolumn{2}{c}{LM parameters}
\\
\cline{2-3}
 & consistent norm. & fixed norm.
\\
\hline
  $\displaystyle\frac{K_\infty}{6}=$
  &
  $\displaystyle T_\mathrm{F}^*\left(1+{F_0}\right)$
  &
  $\displaystyle T_\mathrm{F}\left(\frac{m}{m^*}+{f_0}\right)$
\\[6pt]
  $\displaystyle\frac{m}{m^*}=$
  &
  $\displaystyle\frac{1}{1+\frac{F_1}{3}}$
  &
  $\displaystyle 1-\frac{f_1}{3}$
\\[6pt]
  $\displaystyle 3a_\mathrm{sym}=$
  &
  $\displaystyle T_\mathrm{F}^*\left(1+{F'_0}\right)$
  &
  $\displaystyle T_\mathrm{F}\left(\frac{m}{m^*}+{f'_0}\right)$
\\[6pt]
  $\displaystyle 1\!+\!\kappa_\mathrm{TRK}=$\hspace*{0.3em}
  &
  $\displaystyle\frac{m}{m^*}\left(1+\frac{F'_1}{3}\right)$
  &
  $\displaystyle\frac{m}{m^*}+\frac{f'_1}{3}$
\\
\hline
\end{tabular}
\end{center}
\caption{\label{tab:compare} The two forms of LM parameters
  (\ref{eq:fixed}) and (\ref{eq:consistent}) and their relation to
  nuclear matter parameters (NMP). Upper block: definition of LM
  parameters in terms of NMP. Lower block: NMP computed from LM
  parameters.  The kinetic energies $T_F$ and $T^*_F$ are defined in
  equation (\ref{eq:Tkin}). }
\end{table}

First we look at nuclear matter properties (NMP) which provide a
unique characterization of the basic nuclear response properties in
the volume: incompressibility $K_\infty$, effective mass $m^*/m$,
symmetry energy $a_\mathrm{sym}$, and Thomas-Reiche-Kuhn (TRK) sum
rule enhancement $\kappa_\mathrm{TRK}$.  The first two are isoscalar
response properties and the second two are isovector properties.  The
$\kappa_\mathrm{TRK}$ is a way to parametrize the isovector effective
mass \cite{Bender_2003}.  All four are the response properties in the
excitations channels with natural parity. The NMP for spin modes are
not nearly that well developed, particularly because the data basis on
magnetic excitations is not strong enough to support unambiguous
extrapolation to bulk. Thus we concentrate first on the group of
natural parity modes.  In many of the expressions for NMP appears the
(effective) nucleon mass frequently in a combination which is the
kinetic energy $T_\mathrm{F}$ of bulk matter. To simplify notations,
we introduce for it the abbreviations
\begin{equation}
  T_\mathrm{F}
  =
  \frac{\hbar^2k_\mathrm{F}^2}{2m}
  \quad,\quad
  T_\mathrm{F}^*
  =
  \frac{\hbar^2k_\mathrm{F}^2}{2m*}
  =
  \frac{m}{m^*}T_\mathrm{F}
  \quad.
    \label{eq:Tkin}
\end{equation}
Columns 1 and 2 in the upper block of Table \ref{tab:compare} list the
LM parameters in effective-mass and bare-mass normalization together with their
relations to NMP.  The parameters in effective-mass normalization (column 1)
demonstrate nicely the interplay between mean field (terms with the
leading contribution ``1'')
and the residual interaction (terms with F's). With bare-mass normalization, the
terms representing the mean field are in most cases $m/m^*$ which takes
into account that self-consistent models can stretch or squeeze the level
spacing and the residual interaction thus has to work against the
modified level density, similar to the backflow effect
eq. (\ref{eq:backflow}) for currents.
The lower block shows, in turn, how NMP are computed from LM
parameters.  Again, the place where the effective mass enters makes
the crucial difference between bare-mass normalization and effective-mass
normalization. Particularly noteworthy are the entries for $f_1$ and $f'_1$,
or $F_1$ and $F'_1$ respectively. These show that self-consistent
models establish an intimate connection between these first-order
parameters and effective masses $m/m^*$ and $\kappa_\mathrm{TRK}$. One
is not allowed to change one without consistently modifying the
other. This counterplay is also reflected in the backflow correction
eq. (\ref{eq:backflow}) for flow observables.

The dimensionless LM parameters allow also to express the stability
conditions for excitations modes. These are $F^{(\prime)}_0>-1$ for
effective-mass normalization or $\frac{m^*}{m}{f^{(\prime)}_0}>-1$ for bare-mass
normalization, and similarly $F^{(\prime)}_1>-3$ or
$\frac{m^*}{m}f^{(\prime)}_1>-3$ for $l=1$
(where the compact upper index $(\prime)$ means that this holds
for $F$ as well as for $F'$ type parameters).
The stability conditions look more natural for effective-mass normalization
while one has first to undo the $m/m^*$ factor in case of bare-mass
normalization.

\begin{figure}[h!]
\centerline{\includegraphics*[width=1.2\linewidth,angle=0]{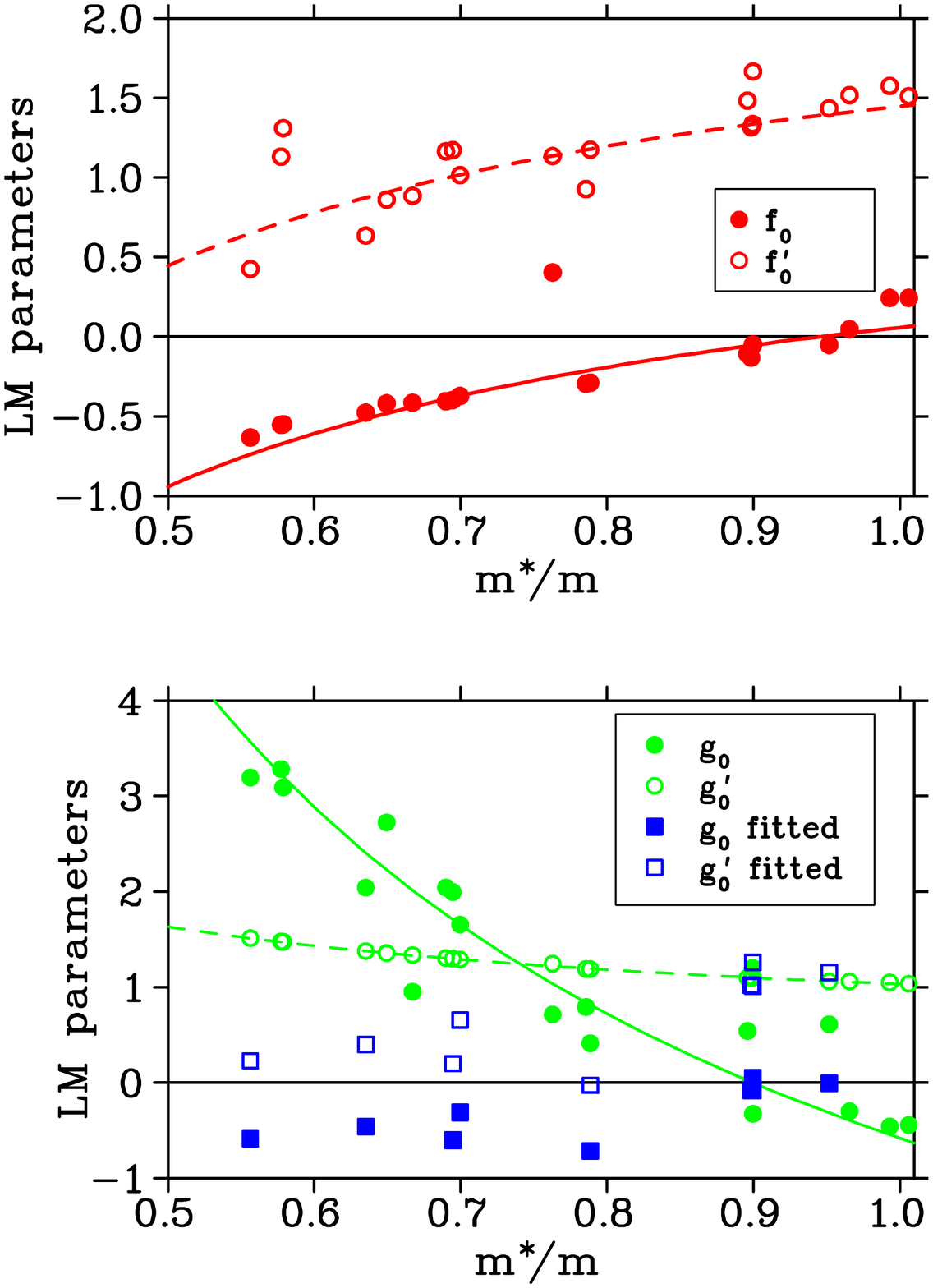}}
\caption{\label{fig:trends-mstar} Dependence of the LM parameters on
  the effective mass $m^*/m$. Upper panel: $f_0$ (filled red circles),
  $f'_0$ (open red circles) derived from the collection of the most
  widely used Skyrme parametrizations given in Table I of
  \cite{Tselyaev_2019}.  The lines indicate the trends $a+bm/m^*$ of
  $f_0$ (solid line) and $f'_0$ (dashed line), computed with the NMP of
  SV-bas (Table \ref{tab:LMp}) except for $m^*/m$ which is
  varied.  Lower panel: $g_0$ (filled green circles) and $g'_0$ (open
  green circles) for the same Skyrme parametrizations. The lines
  indicate again the trends $a+bm/m^*$ (solid line for $g_0$, dashed
  line for $g'_0$). Also shown are the adjusted LM spin parameters
  which reproduce the magnetic dipole states in $^{208}$Pb ($g_0$ as
  filled blue squares, $g'_0$ as open blue squares) for the
  corresponding Skyrme parametrizations taken from
  \cite{Tselyaev_2019}.  }
\end{figure}
As argued above, the parameters $f_0$, $f'_0$, defined with bare-mass
normalization, represent directly the strength of the residual interaction.
The first two lines of Table \ref{tab:compare} show a clear dependence
$f^{(\prime)}_0=c-m/m^*$ where $c$ is some constant: The smaller
$m^*$, the stronger the isoscalar interaction and the weaker the isovector one
which is necessary to
counterweight the lower level density at the Fermi surface (the
``backflow effect'' for the RPA interaction). The upper panel of
Fig. \ref{fig:trends-mstar} shows these trends for the natural-parity channel
together with the values for $f^{(\prime)}_0$ from a representative
set of well working Skyrme parametrizations.
The results from the realistic parametrizations fit nicely to the
analytical trend and so confirm the need to properly counterweight the
level-spreading effect of the effective mass.

\begin{table*}
\begin{tabular}{l|cccc|ccc|ccc}
 EDF & $m^*/m$ & $K_{\infty}$ & $\kappa_{\mbss{TRK}}$ & $a_{\mbsu{sym}}$ &
  ${F_0}$ & ${F'_0}$ & ${F_1}$ &
 $f_0$ & $f'_0$ & $f_1$ \\
  & & (MeV) && (MeV) &  & &  &  & & \\
\hline
 SV-bas    & 0.90 & 233 & 0.4 & 30
 & $-0.05$ & $\hfm$1.20 & $-0.30$
 & $-0.05$ & $\hfm$1.34 & $-0.33$ \\
 SV-sym34  & 0.90 & 234 & 0.4 & 34
 & $-0.04$ & $\hfm$1.50 & $-0.30$
 & $-0.05$ & $\hfm$1.67 & $-0.33$ \\
 SV-mas10  & 1.00 & 234 & 0.4 & 30
 & $\hfm$0.06 & $\hfm$1.45 & $\hfm$0.00
 & $\hfm$0.06 & $\hfm$1.45 & $\hfm$0.00 \\
 SV-mas07  & 0.70 & 234 & 0.4 & 30
 & $-0.26$ & $\hfm$0.71 &  $-0.90$
 & $-0.37$ & $\hfm$1.01 & $-1.29$ \\
 SV-K218   & 0.90 & 218 & 0.4 & 30
 & $-0.12$ & $\hfm$1.18 & $-0.30$
 & $-0.13$ & $\hfm$1.32 & $-0.34$ \\
 SV-m64k6  & 0.64 & 241 & 0.6 & 27
 & $-0.30$ & $\hfm$0.40 & $-1.09$
 & $-0.48$ & $\hfm$0.64 & $-1.72$ \\
 SV-m56k6  & 0.56 & 255 & 0.6 & 27
 & $-0.35$ & $\hfm$0.24 & $-1.33$
 & $-0.63$ & $\hfm$0.43 & $-2.39$ \\
 \end{tabular}
\caption{\label{tab:LMp}
Nuclear matter parameters and spin-independent LM parameters, in both normalizations.}
\end{table*}

For completeness, we show in Table \ref{tab:LMp} the NMP and
corresponding LM parameters for a selection of Skyrme parametrizations
with systematically varied NMP \cite{Kluepfel_2009,Lyutorovich_2012}.
The detailed expressions of the LM parameters in terms of the
parameters of the Skyrme interaction (\ref{eq:skenfun}) are given in
Appendix \ref{app:lmp}.

\subsection{The spin channel}
\label{sec:spinLM}

Now we turn to the spin channel and we will see that the case is
dramatically different. A first problem is that we do not have well
established NMP for spin response and that spin modes in finite nuclei
are not as prominent as giant resonances of natural parity.  Both
together leaves the empirical calibration of the residual interaction
in the spin channel an open problem \cite{Tselyaev_2019}.  Second, in
many mean-field models, the spin channel is determined once the
natural-parity response is fixed.  For example, relativistic
mean-field models tie spin properties and kinetic properties closely
together \cite{Rei89aR}.  This need not to be
beneficial if it turns out that the ``predictions'' thus obtained are
wrong. That is the aspect which we will address here for the case of
the SHF model.

The spin properties in the Skyrme EDF's are not uniquely fixed.  These
leaves different options for its choice \cite{Pot10a} which lead to
rather different result for the LM parameters of $G$ type:
\begin{enumerate}
  \item\label{it:full} One can understand SHF as stemming from the
    effective density-dependent zero-range interaction
    (\ref{eq:skenfun}) which determines all spin terms from the given
    natural-parity partners and its NMP.
By combining the formulas of Appendices \ref{app:nmp} and \ref{app:lmp}
this yields
   \begin{subequations}
   \begin{eqnarray}
     g_0\!+\!g_1
     &=&
      -(3\!+\!3\kappa_\mathrm{TRK})
      -\frac{3a_\mathrm{sym}-2\frac{B}{A}}{T_\mathrm{F}}
     +\frac{26}{5}\,\frac{m}{m^*}
     \;,
   \\
     g'_0\!+\!g'_1
     &=&
     \frac{B}{A}\,\frac{1}{T_\mathrm{F}}
     +\frac{3}{5}\,\frac{m}{m^*}
     \;.
   \end{eqnarray}
   \end{subequations}

  \item\label{it:noJ}
    Even when taking the viewpoint of option \ref{it:full}, most
    actual parametrizations drop the tensor spin-orbit terms ``tensor
    terms'' $\propto\vec{J}^2$ which are generated as partners of the
    kinetic terms in the force definition of the SHF functional. This
    yields the variant
   \begin{subequations}
   \label{eq:fullspin-noJ2}
   \begin{eqnarray}
     g_0
     &=&
      -(3+3\kappa_\mathrm{TRK})
      -\frac{3a_\mathrm{sym}-2\frac{B}{A}}{T_\mathrm{F}}
     +\frac{26}{5}\,\frac{m}{m^*}
     \;,
   \\
     g'_0
     &=&
     \frac{B}{A}\,\frac{1}{T_\mathrm{F}}
     +\frac{3}{5}\,\frac{m}{m^*}
     \;,
\label{eq:g0p2}
   \\
     g_1&=& 0\;,
   \\
     g'_1&=& 0\;.
   \end{eqnarray}
   \end{subequations}

  \item\label{it:free}
    One can dismiss the concept of a force and start from an
    energy-density functional in which case the spin terms are
    constrained only by the requirement of Galilean invariance
    leaving a couple of terms open. These can be adjusted
    independently from the terms of natural parity and so allow
    for more flexible tuning of magnetic modes. This has been
    done, e.g., in \cite{Tselyaev_2019}. No closed formula for the $G$
    parameters can be given here.

  \item\label{it:minGal}
    As in option \ref{it:free}, one can start from a Skyrme
    energy-density functional, but now freeze the spin terms by the
    requirement of ``minimal Galilean invariance'' which means to
    discard all spin terms which are not fixed by Galilean invariance
    \cite{Pot10a}. This yields for the $G$ parameters the
    trivial result
    \begin{equation}
       G_0=0\;,\;G'_0=0\;,\;G_1=0\;,\;G'_1=0\;.
    \end{equation}

\end{enumerate}
Let us first investigate the option \ref{it:noJ} which assumes an
underlying Skyrme force and thus predicts the properties in the spin
channel from the known properties in the natural-parity channel.  The
LM parameters are thus given by eqs.  (\ref{eq:fullspin-noJ2}).  The
trend with $m^*$ is of the form
$g^{(\prime)}_0=a^{(\prime)}+b^{(\prime)}m/m^*$ where $a^{(\prime)}$
and $b^{(\prime)}$ are some constant. This looks similar to the trend
for the $f^{(\prime)}_0$. The crucial difference is, however, that the
mass dependence comes with a plus sign. The trend is visualized in the
lower panel of Fig. \ref{fig:trends-mstar}.
Note that the deviation of the open green circles from the dashed line
is negligible because the first term in the right-hand side of Eq. (\ref{eq:g0p2})
is practically the same for all parametrizations.
We see that the $g^{(\prime)}_0$ parameters
increase with decreasing $m^*/m$ which goes the wrong way because it
is counter-productive for compensating the decrease of level density
in the single-particle spectrum. The options \ref{it:full} and
\ref{it:noJ} which understand the SHF model as an effective
interaction is thus to be discarded for principle reasons.

This result has also been found at several places from studying
magnetic excitations in finite nuclei, see
e.g. \cite{Nesterenko_2010,Tselyaev_2019}. In \cite{Tselyaev_2019},
the spin-parameters in the Skyrme functional had been adjusted freely
to M1 modes in finite nuclei. This corresponds to option \ref{it:free}
in the above list. The resulting $g^{(\prime)}$ are shown as squares
in Fig. \ref{fig:trends-mstar}. The $g_0$ stay close to zero
for the parametrizations with $m^*/m \approx 1$.
The $g'_0$ a bit larger, still being small. Both show a slight tendency to
decrease with decreasing $m^*/m$ which is the expected trend.
This empirical result allows also the option \ref{it:minGal} for
$g_0$. This is  not so clear for $g'_0$. To be on the safe side, the
option  option \ref{it:free} turns out to be the recommended option.

\section{LM parameters and resonance excitations in $^{208}$Pb}
\label{sec:electric}

\subsection{The random phase approximation (RPA)}

In this section, we are going to investigate excitation properties in
a finite nucleus, namely $^{208}$Pb. The most often used method for
calculating excitation properties in nuclear physics is the RPA and
its various extended versions.  There exists numerous different
derivations which all lead to the same basic RPA equation
\cite{Ring80}:
\bea
\label{eq:1}
  && \left( \epsilon_{\nu_1}-\epsilon_{\nu_2} - \Omega_m\right)\chi^{(m)}_{\nu_1 \nu_2}
  \nonumber\\
  &=& (n_{\nu_1} - n_{\nu_2})
  \sum_{\nu_3 \nu_4}F^\mathrm{ph}_{\nu_1 \nu_4 \nu_2 \nu_3} \chi^{(m)}_{\nu_3 \nu_4}
  \;.
\eea
The $\chi^{(m)}_{\nu_1 \nu_2}$ are the ph excitation amplitudes
in the single-particle configuration space, $F^{ph}$ is the
$ph$-interaction, $\epsilon_{\nu}$ are the sp energies and $\Omega_m$
the excitation energies of the nucleus. There exist two different
methods to determine the input data:

(I) The phenomenological shell model approach where one starts with an
empirically adjusted single particle model and parametrizes the
$ph$-interaction. A very successful approach in this connection is the
Landau-Migdal theory \cite{Migdal_1967,Speth_1991}.

(II) The self-consistent approach where one starts with an
effective energy-density functional (EDF) from which one derives
the single-particle quantities as well as the $ph$-interaction \cite{Rei92b}.
In this paper we discuss particularly Skyrme type
EDF's.

There exist various extended versions of RPA which include
configurations beyond $1ph$, e.g., phonon coupling in time blocking
approximation (TBA), for details see
\cite{Tselyaev_2007,Tselyaev_2016}.  Most of these models employ again
the basic ph interaction $F^\mathrm{ph}$. Thus no new parameters have
to be introduced.

The RPA equation (\ref{eq:1}) shows that there are two basic
ingredients which determine at the end the excitation spectra: the $1ph$
energies $\epsilon_{p}-\epsilon_{h}$ and the $ph$-interaction
$F^\mathrm{ph}$. The energies are determined with the ground state
which leaves little leeway for tuning.  The $ph$-interaction is
exclusively seen in the excitations and most of their impact can be
characterized in simple terms through the LM parameters as done
throughout this paper.

\subsection{Giant resonances}

We start with excitations of natural parity, also called electrical
modes.  Their spectral distribution in a heavy nucleus as $^{208}$Pb
and in channels with low angular momentum $L$ is dominated by one
strong peak called a giant resonance.  Most prominent are the
isoscalar giant monopole resonance (GMR), the isoscalar giant
quadrupole resonance (GQR), and the the isovector giant dipole
resonance (GDR). All three resonances can be characterized by
one number, the resonance energy, which we will use now for
the looking at trends and relations to LM parameters.
\begin{figure}
\centerline{\includegraphics[width=0.7\linewidth,angle=0]{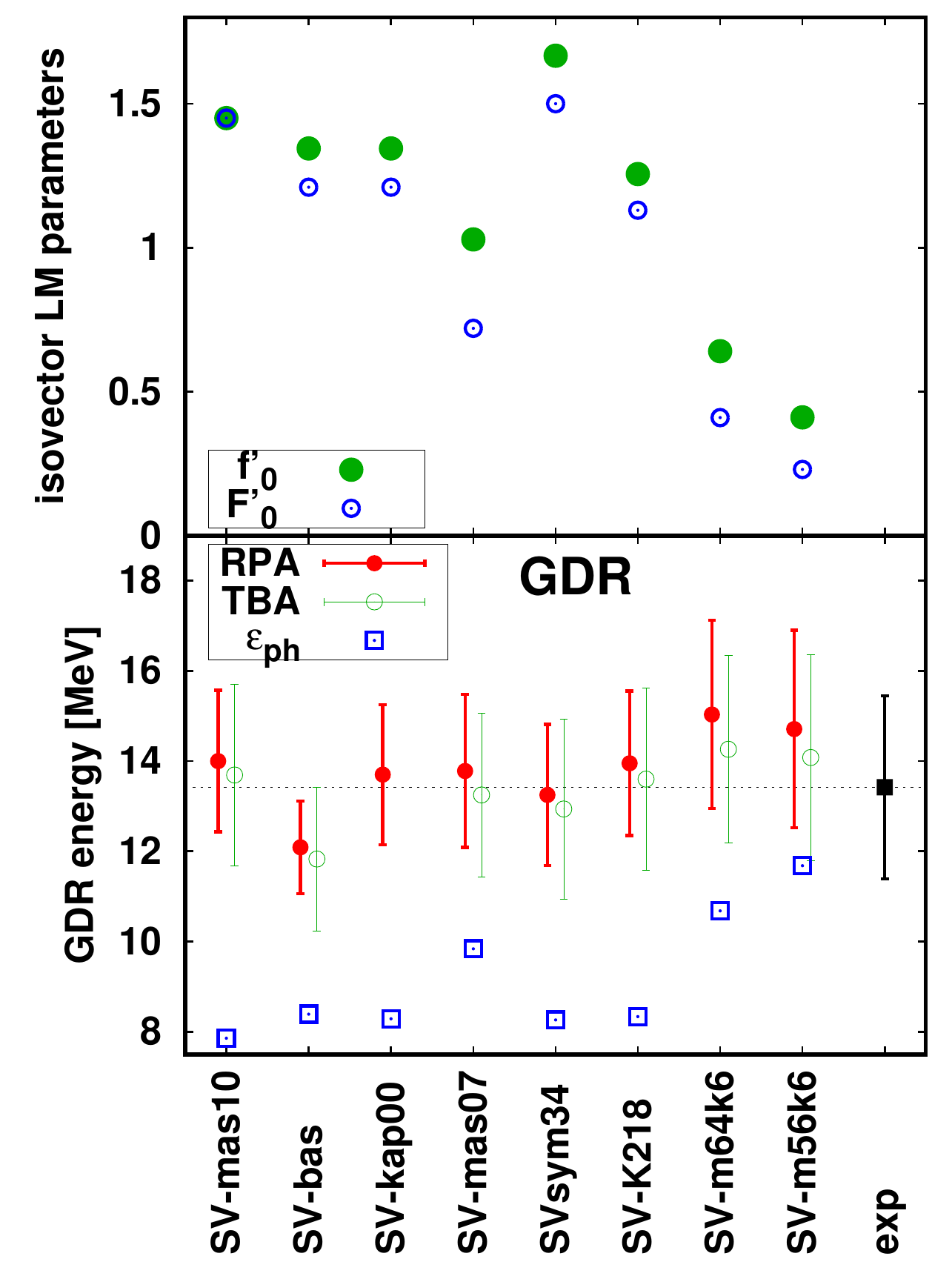}}
\centerline{\includegraphics[width=0.7\linewidth,angle=0]{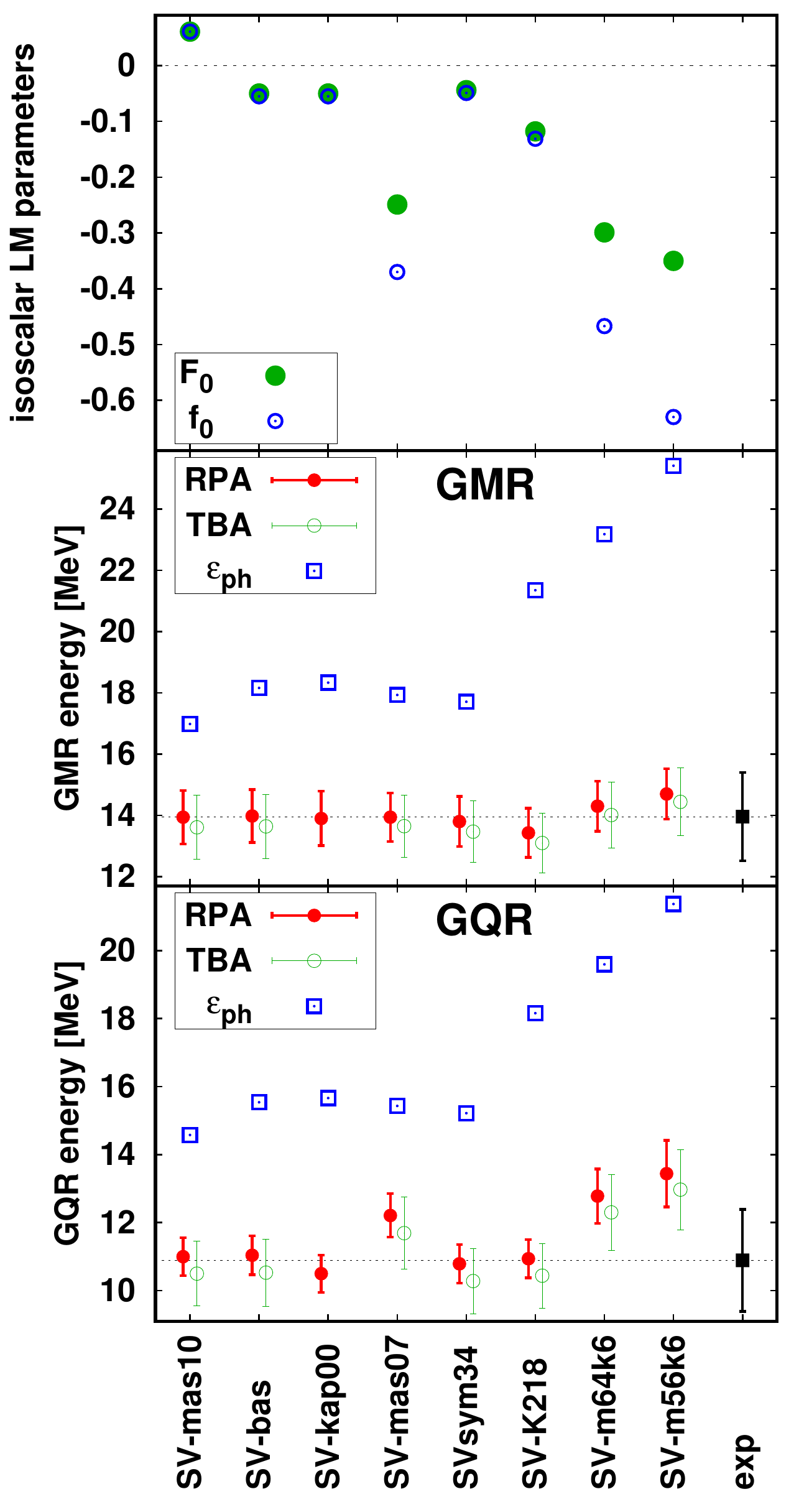}}
\caption{\label{fig:js2} Collection of giant-resonance properties in
  $^{208}$Pb together with LM parameters for a representative set of
  Skyrme parametrizations covering a variation of all four NMP
  \cite{Kluepfel_2009,Lyutorovich_2012}.  Upper block: isovector
  properties, LM parameters in upper panel and RPA properties
  (resonance energies, average $1ph$ energies) in the lower panel.
  Lower block: isoscalar properties, LM parameters in upper panel, RPA
  properties in middle and lower panel. In addition to RPA results,
  also results from TBA are shown. }
\end{figure}
Fig. \ref{fig:js2} collects giant-resonance properties together with
the leading LM parameters (upper panels) for a variety of Skyrme
parametrizations with systematically varied NMP, see Table~\ref{tab:LMp}.
In order to check the impact of complex configurations
beyond RPA, we compare resonance energies from RPA with those from
TBA. The differences in the energies are small while the resonance
width is significantly affected by the complex configurations in TBA
\cite{Tselyaev_2017,Tselyaev_2018,Tselyaev_2019}. At present, we focus
on resonance energies and can ignore the small difference between RPA
and TBA. Together with the resonance energies, we shows also the
average $1ph$ energies $\varepsilon_{ph}$ (averaged over the $1ph$
spectrum weighted by the transition operator of each mode).  The
difference between $\varepsilon_{ph}$ and the resonance energy
visualizes the impact of the $ph$-interaction and that is obviously
considerable, strongly attractive in the isoscalar modes
(lower block)
and strongly repulsive in the isovector modes
(upper block).
Note that the Skyrme parametrizations are sorted in order of decreasing
effective mass with $m^*/m=1$ to the left and the lowest $m^*/m=0.56$
to the right. It is obvious that the $\varepsilon_{ph}$ increase while
the resonance energies change comparatively little. The increase of
$\varepsilon_{ph}$ is largely compensated by a properly counter-acting
trend of the $ph$-interaction.  This trend can be nicely read off from
the LM parameters shown in the upper panels. It is the same as
shown already in Fig. \ref{fig:trends-mstar} and we learn from the
present figure that the trend $\propto c-m/m^*$ which is typical
for the LM parameters in the  natural parity channels is exactly
what is needed to compensate the dilution of $1ph$ spectra with
decreasing effective mass.

Although the variations of resonance energies are small as compared to
the effects of the $ph$-interaction, there is important systematics in
it. They demonstrate the known intimate connection between NMP and
resonance energies \cite{Kluepfel_2009}: the giant monopole resonance
(GMR) is related exclusively to the incompressibility $K_{\infty}$,
the giant dipole resonance (GDR) to the sum rule enhancement
$\kappa_\mathrm{TRK}$, and the giant quadrupole resonance (GQR) to the
effective mass $m^*/m$.  These trends are much more subtle than the
dramatic trends for the $\varepsilon_{ph}$.  It is remarkable how the
interplay between mean-field and $ph$-interaction can recover the
subtle trends.

\subsection{The magnetic case}
\label{sec:mag}

In case of magnetic modes, there are no isoscalar spin dependent
resonances known which suggests that the spin-dependent isoscalar
$ph$-interaction is weak.  On the other hand, there exist collective
neutron-particle proton-hole resonances in nuclei with neutron
excess. The best known resonances are the ($1^+$) Gamow-Teller
resonances.  The corresponding unperturbed $1ph$-strength is shifted
to higher energies which is a clear indication that the spin-isospin
$ph$-interaction has to be strongly repulsive which was confirmed in
Ref. \cite{Wakasa_2012} comparing the experimental GT resonance in
$^{208}$Pb together with two theoretical results. However, the
Gamow-Teller resonances reside in a regime where effective energy
functionals are most probably insufficient. We thus concentrate on the
low-energy M1 modes.

The $1^+$-states in $^{208}$Pb are a nice example for the behavior of
the spin-depended isoscalar and isovector interaction.  There is an
isoscalar state at $E_1$~=~5.84 MeV which is close to the uncorrelated
proton and neutron spin-orbit doublets ${\epsilon}^{\pi}_{ph}$~=~5.55
MeV and ${\epsilon}^{\nu}_{ph}$~=~5.84 MeV and a couple of isovector
$1^+$ states with the mean energy $E_2$~=~7.39 MeV. This again shows
that the spin-dependent isoscalar $ph$-interaction is weak and the
spin-dependent isovector $ph$-interaction is strongly repulsive. In a
recent publication by our group \cite{Tselyaev_2019} we investigated
these $1^+$ states in the framework of RPA using various Skyrme
parameter sets with different effective masses. There we took the Skyrme
functional as derived from a Skyrme interaction with all spin terms
fixed by the model, option \ref{it:noJ} of section \ref{sec:spinLM}.
\begin{figure}
\centerline{\includegraphics[width=0.7\linewidth,angle=0]{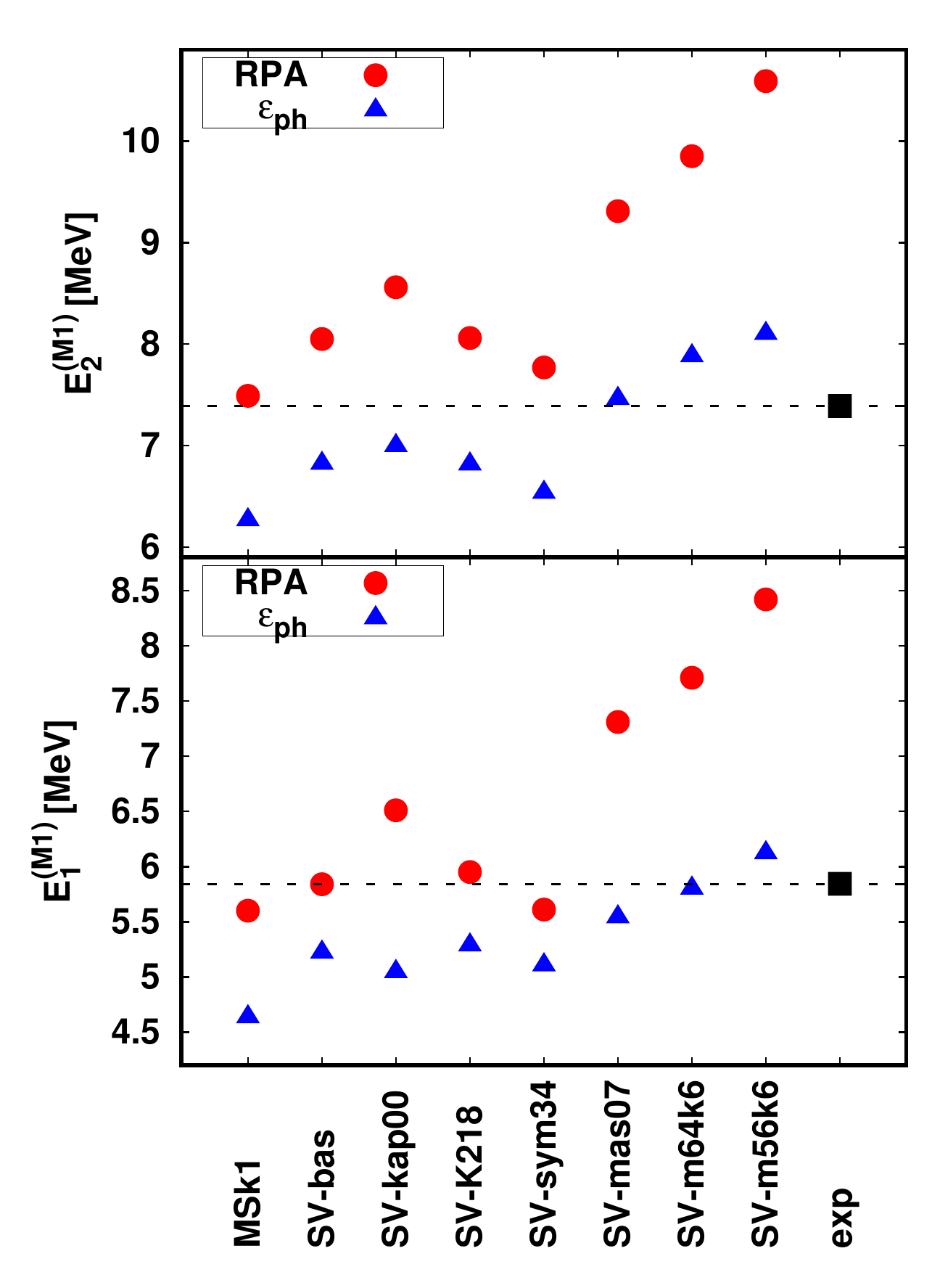}}
\caption{\label{fig:M1-compare} Energies of the lower ($E^{(M1)}_1$; lower figure) and
  higher ($E^{(M1)}_2$; upper figure) M1 states in $^{208}$Pb calculated within RPA for a
  selection of different Skyrme parametrizations (full red circle)
  compared with the experimental values (black box and faint dashed
  line). Also shown are the energy of the unperturbed
  $\varepsilon_{ph}$ from proton spin-orbit pair (lower figure) and neutron
  spin orbit pair (upper figure), indicated by blue triangles.}
\end{figure}
Fig. \ref{fig:M1-compare} shows the RPA results of the isoscalar and
isovector M1 modes together with the unperturbed $1ph$ energies.  The
trends of the $1ph$ energies are the same as for the giant resonances
in Fig. \ref{fig:js2} and the computed M1 energies amplify this trend
driving the RPA results far off the experimental values. This
demonstrates on the grounds of the empirical results that the option
to take the Skyrme interaction literally is inappropriate for
magnetic modes.  In the paper \cite{Tselyaev_2019}, we had also
considered the spin terms in the Skyrme functional as free for
independent calibration (option \ref{it:free} in section
\ref{sec:spinLM}). The energies of the M1 modes reproduce, by
construction, the experimental energies and are thus not shown in the
figure. The non-trivial message in this respect is that one can do
such fine tuning of spin modes without destroying the overall quality
of the parametrization.

\section{Conclusion}
\label{sec:sum}

We explored in this paper the general pattern and trends of
Landau-Migdal (LM) parameters in connection with the self-consistent
models using the Skyrme energy-density functional (EDF). As starting
point, we reviewed the channel of natural-parity excitations. The well
known experience is that giant resonances are well described for
several Skyrme EDF's although they can have quite different effective
nucleon masses. This is surprising because changing the effective mass
changes energy spacing of particle-hole ($ph$) states
dramatically. The fact that giant resonances do not change that much
implies that the change in $ph$ spacing is compensated by a
corresponding change in the residual $ph$-interaction: smaller
effective mass gives larger $ph$ spacings and thus the $ph$-interaction
has to be more attractive (isoscalar channel) or less
repulsive (isovector channel). In the present paper, we studied in
quantitative detail these correlations between $ph$-spacing and strength
of residual $ph$-interaction for Skyrme EDF's. The latter were
quantified in terms of LM parameters which depend, apart from $m^*/m$,
only on five nuclear matter parameters (Fermi momentum $k_F$, bulk
binding energy $B/A$, incompressibility $K_{\infty}$, symmetry energy
$a_{\mbsu{sym}}$ and Thomas-Reiche-Kuhn sum rule enhancement
$\kappa_{\mbss{TRK}}$). Modern Skyrme parametrizations have only a
moderate dispersion in theses parameters leaving close correlations
between $m^*/m$ and LM parameters. As expected, with decreasing
effective $m^*/m$ the LM parameter $f_0$ (isoscalar) becomes more
attractive and $f'_0$ (isovector) less repulsive and the trend has
exactly the right amount to guarantee that the energies of isoscalar
and isovector electrical giant resonance are well reproduced by Skyrme
parametrizations with different effective masses.

Then we turned to magnetic modes, i.e. excitations with unnatural
parity.  The situation is found to be completely different.  First of
all, there exist no well settled magnetic bulk properties which may be
included in the fitting of the EDF parameters. This leaves several
options for determining the EDF in the spin channel. Either one derives
the spin parameters from the zero-range Skyrme force as done
traditionally, or one dismisses all terms which are not required by
Galilean invariance, or takes spin-sensitive data to calibrate them.
Second, there are no strong collective magnetic resonances known with
the exception of the GT-resonances in neutron rich nuclei which,
however, is likely to lie outside the range of a description by Skyrme
EDF's. Thus we take as reference here the strongest isoscalar and
isovector M1-states in $^{208}$Pb. The isoscalar state is close to the
two (experimental) spin-orbit partners while the more fragmented
isovector states are shifted by about two MeV to higher energies.
Taking the definition of spin parameters in the EDF from the Skyrme
force runs into difficulties for M1 resonances in $^{208}$Pb. The RPA
results do not describe the data and there do not exist the clear
correlations between unperturbed $ph$ states and RPA results.  The
main point of our paper is that this problem is already apparent from
bulk properties, namely looking at the trends of the spin dependent LM
parameters $g_0$ and $g'_0$ as function of $m^*/m$. These trends are
going into the opposite direction as the well performing LM parameters
$f_0$ and $f'_0$ in the natural-parity channel. This provides, already at the
level of bulk properties, a strong argument against the definition of
a Skyrme EDF by a Skyrme force. The argument is corroborated by the
observation that the values of $g_0$ and $g'_0$ differ
substantially from those obtained previously by a fit to the empirical
M1 resonances. This altogether demonstrates once again that the
spin channel in Skyrme EDF's is different  and still require careful
calibration.

\begin{acknowledgements}
This research was carried out using computational resources provided
by the Computer Center of St. Petersburg State University.
\end{acknowledgements}

\appendix

\section{On the density dependence of LM parameters}
\label{sec:densdepLM}

As said above, the LM theory for finite nuclei as well as the SHF model
augment the LM parameters with some density dependence.
\begin{figure}
\begin{center}
\includegraphics*[height=7.0cm,angle=0]{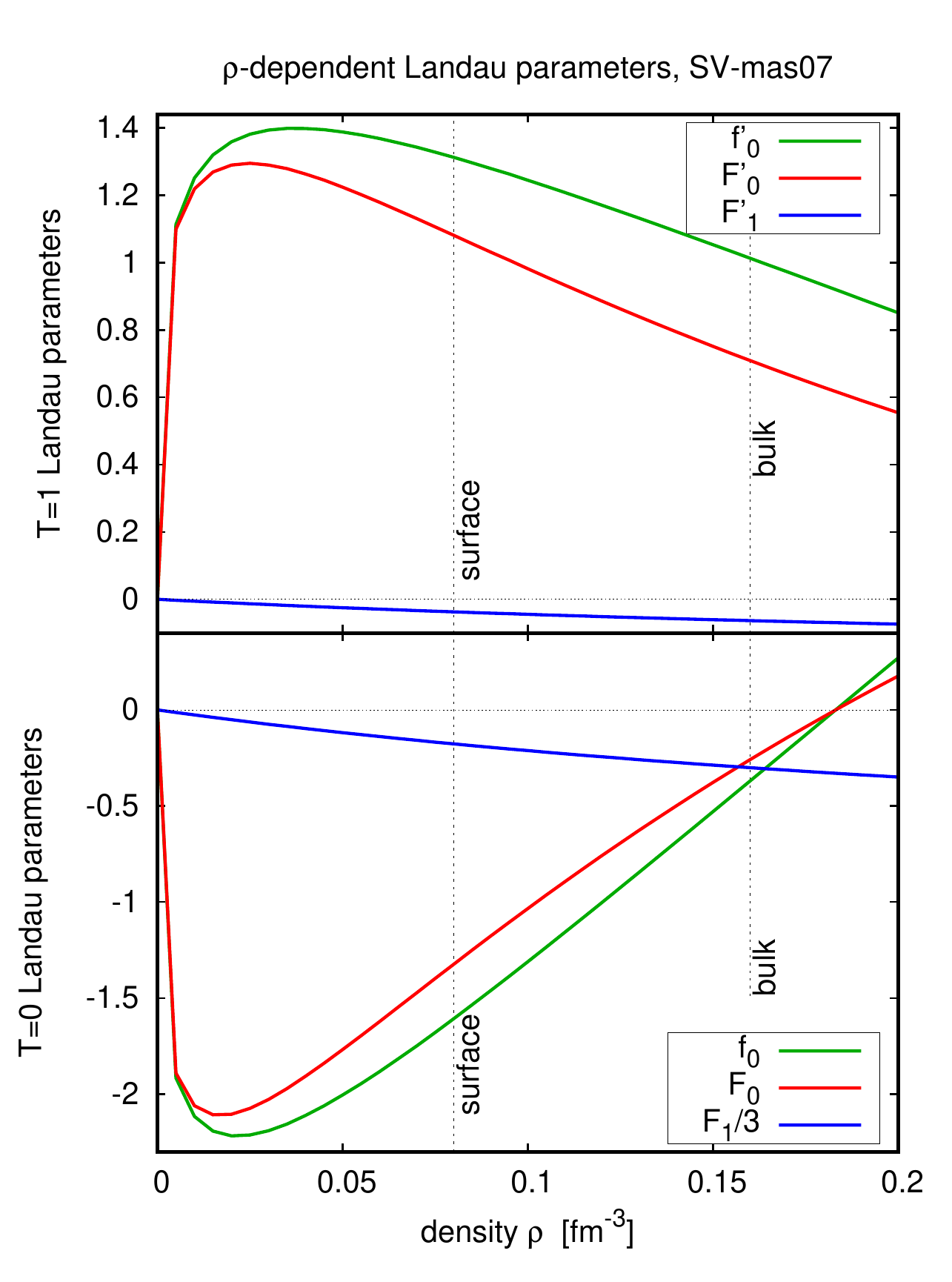}
\end{center}
\caption{\label{fig:js1}
Density dependence of the isovector LM parameters $f'_0$, $F'_0$, $F'_1$
in the upper part and the isoscalar LM parameters $f_0$, $F_0$, $F_1$ in the lower part.
The quantities are derived from the Skyrme parametrization SV-mas07.}
\end{figure}
Fig. \ref{fig:js1} shows the density dependent LM parameters for the
parametrization SV-mas07. Near bulk density, it is linear similar to
LM theory.  But it differs dramatically from linear behavior at low
densities.

\section{Low-lying collective electric states}
\label{sec:low}

\begin{figure}
\begin{center}
\centerline{\includegraphics[width=\linewidth,angle=0]{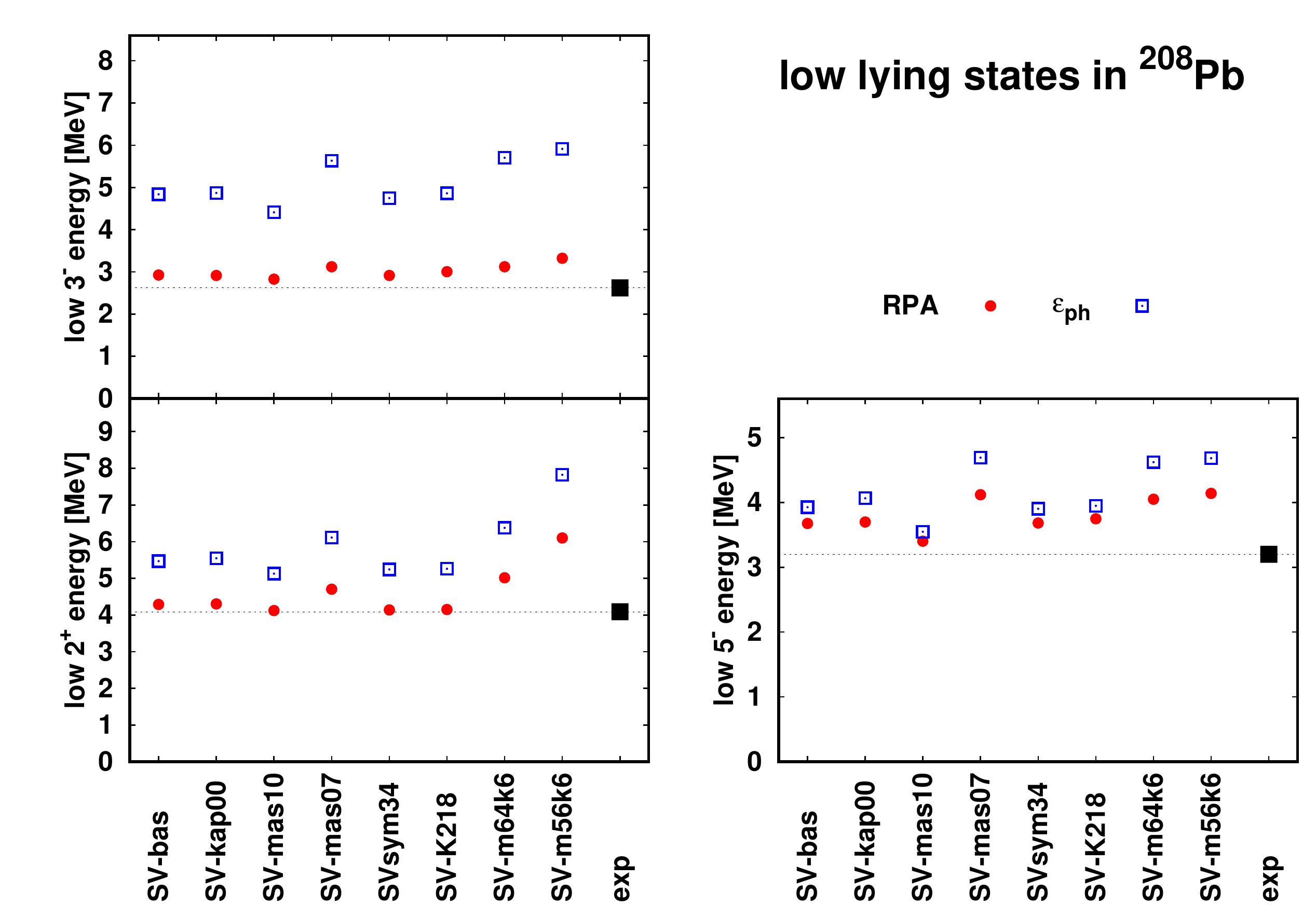}}
\end{center}
\caption{\label{fig:low-compare} Energies of the first excited  $3^-$,~$5^-$
and~$2^+$ states in $^{208}$Pb calculated within RPA. We compare the results of different
Skyrme parametrizations with the data. We also show the energy of the lowest unperturbed
$1ph$-pair for each multipolarity indicated by blue quads.}
\end{figure}

In Fig. \ref{fig:low-compare} we present the energies of the first
$3^-$,~$5^-$~and $2^+$ states in $^{208}$Pb calculated in RPA with
various Skyrme parametrizations. For each parameter set also the
energy of the lowest unperturbed $1ph$ state in the corresponding
channel is given. From numberless calculations, e.g. Ref. \cite{Ring_1973plb},
we know that the lowest
$3^-$ state is the most collective state in
$^{208}$Pb. Many $1ph$ state within the $1\hbar
\omega$-shell contribute coherently which gives rise to the well known
large transition probability and large energy shift. This is nicely
demonstrated in the left upper section of Fig. \ref{fig:low-compare}
where all $1ph$ energies stay far above the finally lowest state (red
dots).  To the $5^-$ state also many $1ph$ states within the $1\hbar \omega$-shell
contribute but obviously in this case the $ph$-interaction is too weak to generate
a strongly collective state. Therefore the shift from the unperturbed states
is much smaller and reaches in no case the experimental line. For the $2^+$ states
only two neutron and two proton $1ph$ states within the $1\hbar \omega$-shell contribute.
On the other hand many $1ph$ states from the $1\hbar \omega$-shell contribute and give
rise to a relatively large transition moment. The energy shifts are smaller than
in the $3^-$ case but reaches in most cases the experimental value. The down-shift
of the energy comes along with an enhanced transition moment (not shown here) which is
another realization of collectivity (coherent superposition of many
$1ph$ states). Most collective resonance in that respect is the $3^-$ state and it is no
surprise that we see, again, the same feature as for the giant
resonances, namely that the uncoupled $1ph$ energies change with Skyrme
force while the RPA results are practically the same. From this we conclude that
for collective states the \emph{back-flow}
is an important corrective mechanism.

\section{Nuclear matter properties}
\label{app:nmp}
Within the density functional theory,
the properties of symmetric infinite nuclear matter
(the Fermi momentum $k_{\mbsu{F}}$,
the total binding energy per nucleon $B/A$,
the nuclear matter incompressibility $K_{\infty}$,
the symmetry energy $a_{\mbsu{sym}}$,
the enhancement factor of the Thomas-Reiche-Kuhn sum rule $\kappa_{\mbss{TRK}}$,
and the effective mass $m^*$)
are determined by the parameters of the energy-density functional.
In the case of Skyrme EDF (\ref{eq:skenfun}) the respective equations
have the following form (see, e.g., Ref. \cite{Chabanat_1997})
\bea
0 &=& \frac{2}{5} T_{\mbsu{F}} + \frac{3}{8} t_0\rho_{\mbsu{eq}}
+ \frac{1}{16} t_3 (\alpha + 1) \rho_{\mbsu{eq}}^{\alpha + 1}
\nonumber\\
&+& \frac{1}{16} \Theta_{\mbsu{s}} k^2_{\mbsu{F}} \rho_{\mbsu{eq}}\,,
\label{nmp:eql}\\
- B/A &=& \frac{3}{5} T_{\mbsu{F}} + \frac{3}{8} t_0\rho_{\mbsu{eq}}
+ \frac{1}{16} t_3 \rho_{\mbsu{eq}}^{\alpha + 1}
\nonumber\\
&+& \frac{3}{80} \Theta_{\mbsu{s}} k^2_{\mbsu{F}} \rho_{\mbsu{eq}}\,,
\label{nmp:ba}\\
K_{\infty} &=& \frac{6}{5} T_{\mbsu{F}} + \frac{9}{4} t_0\rho_{\mbsu{eq}}
+ \frac{3}{16} t_3 (\alpha + 1) (3\alpha + 2) \rho_{\mbsu{eq}}^{\alpha + 1}
\nonumber\\
&+& \frac{3}{4} \Theta_{\mbsu{s}} k^2_{\mbsu{F}} \rho_{\mbsu{eq}}\,,
\label{nmp:knm}\\
a_{\mbsu{sym}} &=& \frac{1}{3} T_{\mbsu{F}} - \frac{1}{8} t_0 (2x_0 + 1) \rho_{\mbsu{eq}}
- \frac{1}{48} t_3 (2x_3 + 1) \rho_{\mbsu{eq}}^{\alpha + 1}
\nonumber\\
&+& \frac{1}{24} (2\Theta_{\mbsu{s}} - 3\Theta_{\mbsu{v}}) k^2_{\mbsu{F}} \rho_{\mbsu{eq}}\,,
\label{nmp:as}\\
\kappa_{\mbss{TRK}} &=& \frac{m \rho_{\mbsu{eq}}}{4 \hbar^2} \Theta_{\mbsu{v}}\,,
\label{nmp:kap}\\
\frac{m}{m^*} &=& 1 + \frac{m \rho_{\mbsu{eq}}}{8 \hbar^2} \Theta_{\mbsu{s}}\,,
\label{nmp:meff}
\eea
where $\rho_{\mbsu{eq}} = 2k_{\mbsu{F}}^3/3\pi^2$ is the equilibrium density,
$T_{\mbsu{F}} = \hbar^2 k_{\mbsu{F}}^2/2m$,
\bea
\Theta_{\mbsu{s}} &=& 3t_1 + (5 +4x_2)t_2\,,
\label{def:thts}\\
\Theta_{\mbsu{v}} &=& (2 + x_1)t_1 + (2 + x_2)t_2\,.
\label{def:thtv}
\eea

\section{Landau-Migdal parameters}
\label{app:lmp}

The Landau-Migdal parameters deduced from the Skyrme EDF (\ref{eq:skenfun})
are related with the parameters of this functional by the formulas
(see, e.g., Refs.~\cite{vanGiai_1981,Margueron_2002})
%
\bea
C_0^* F^{\vphu}_0  &=& \frac{3}{4} t_0
+ \frac{1}{16} t_3 (\alpha + 1) (\alpha + 2) \rho_{\mbsu{eq}}^{\alpha}
\nonumber\\
&+& \frac{1}{8} k^2_{\mbsu{F}}[3t_1 + (5 +4x_2)t_2] \,,
\label{lmp:f01}\\
C_0^* F'_0 &=& -\frac{1}{4} t_0 (1 + 2x_0)
- \frac{1}{24} t_3 (1 + 2x_3) \rho_{\mbsu{eq}}^{\alpha}
\nonumber\\
&+&\frac{1}{8} k^2_{\mbsu{F}}[(1+2x_2)t_2-(1+2x_1)t_1]\,,
\label{lmp:fp01}\\
C_0^* F^{\vphu}_1 &=& -\frac{1}{8} k^2_{\mbsu{F}}[3t_1 + (5 +4x_2)t_2] \,,
\label{lmp:f1}\\
C_0^* F'_1 &=& -\frac{1}{8} k^2_{\mbsu{F}}[(1+2x_2)t_2-(1+2x_1)t_1]\,,
\label{lmp:fp1}
\eea
\bea
C_0^* (G^{\vphu}_0 + G^{\vphu}_1) &=& - \frac{1}{4} t_0 (1 - 2x_0)
\nonumber\\
&-& \frac{1}{24} t_3 (1 - 2x_3) \rho_{\mbsu{eq}}^{\alpha}\,,
\label{lmp:g01}\\
C_0^* (G'_0 + G'_1) &=& - \frac{1}{4} t_0
- \frac{1}{24} t_3 \rho_{\mbsu{eq}}^{\alpha}\,,
\label{lmp:gp01}\\
C_0^* G^{\vphu}_1 &=& \frac{1}{8} [(1 - 2 x_1) t_1 - (1 + 2 x_2) t_2]
k^2_{\mbsu{F}}\,,\qquad
\label{lmp:g1}\\
C_0^* G'_1 &=& \frac{1}{8} (t_1 - t_2) k^2_{\mbsu{F}}\,,
\label{lmp:gp1}
\eea
with $C_0^*$ as defined in Eq. (\ref{eq:C0star}).
Note that $G^{\vphu}_1 = G'_1 = 0$ independently of Eqs.
(\ref{lmp:g1}) and (\ref{lmp:gp1}) for the Skyrme EDFs in which
the $\bfbj^2$ terms are omitted (see Ref.~\cite{Lesinski_2007}
for more detail).

\bibliographystyle{apsrev4-1}
\bibliography{TTT}

\begin{thebibliography}{39}%
\makeatletter
\providecommand \@ifxundefined [1]{%
 \@ifx{#1\undefined}
}%
\providecommand \@ifnum [1]{%
 \ifnum #1\expandafter \@firstoftwo
 \else \expandafter \@secondoftwo
 \fi
}%
\providecommand \@ifx [1]{%
 \ifx #1\expandafter \@firstoftwo
 \else \expandafter \@secondoftwo
 \fi
}%
\providecommand \natexlab [1]{#1}%
\providecommand \enquote  [1]{``#1''}%
\providecommand \bibnamefont  [1]{#1}%
\providecommand \bibfnamefont [1]{#1}%
\providecommand \citenamefont [1]{#1}%
\providecommand \href@noop [0]{\@secondoftwo}%
\providecommand \href [0]{\begingroup \@sanitize@url \@href}%
\providecommand \@href[1]{\@@startlink{#1}\@@href}%
\providecommand \@@href[1]{\endgroup#1\@@endlink}%
\providecommand \@sanitize@url [0]{\catcode `\\12\catcode `\$12\catcode
  `\&12\catcode `\#12\catcode `\^12\catcode `\_12\catcode `\%12\relax}%
\providecommand \@@startlink[1]{}%
\providecommand \@@endlink[0]{}%
\providecommand \url  [0]{\begingroup\@sanitize@url \@url }%
\providecommand \@url [1]{\endgroup\@href {#1}{\urlprefix }}%
\providecommand \urlprefix  [0]{URL }%
\providecommand \Eprint [0]{\href }%
\providecommand \doibase [0]{http://dx.doi.org/}%
\providecommand \selectlanguage [0]{\@gobble}%
\providecommand \bibinfo  [0]{\@secondoftwo}%
\providecommand \bibfield  [0]{\@secondoftwo}%
\providecommand \translation [1]{[#1]}%
\providecommand \BibitemOpen [0]{}%
\providecommand \bibitemStop [0]{}%
\providecommand \bibitemNoStop [0]{.\EOS\space}%
\providecommand \EOS [0]{\spacefactor3000\relax}%
\providecommand \BibitemShut  [1]{\csname bibitem#1\endcsname}%
\let\auto@bib@innerbib\@empty
\bibitem [{\citenamefont {Migdal}(1967)}]{Migdal_1967}%
  \BibitemOpen
  \bibfield  {author} {\bibinfo {author} {\bibfnamefont {A.~B.}\ \bibnamefont
  {Migdal}},\ }\href@noop {} {\emph {\bibinfo {title} {Theory of Finite Fermi
  Systems and Application to Atomic Nuclei}}}\ (\bibinfo  {publisher} {Wiley},\
  \bibinfo {address} {New York},\ \bibinfo {year} {1967})\BibitemShut {NoStop}%
\bibitem [{\citenamefont {Landau}\ \emph {et~al.}(1980)\citenamefont {Landau},
  \citenamefont {Lifshitz},\ and\ \citenamefont {Pitajevski}}]{LanLif9}%
  \BibitemOpen
  \bibfield  {author} {\bibinfo {author} {\bibfnamefont {L.~D.}\ \bibnamefont
  {Landau}}, \bibinfo {author} {\bibfnamefont {E.~M.}\ \bibnamefont
  {Lifshitz}}, \ and\ \bibinfo {author} {\bibfnamefont {L.~P.}\ \bibnamefont
  {Pitajevski}},\ }\enquote {\bibinfo {title} {{C}ourse of {T}heoretical
  {P}hysics 9--{S}tatisical {P}hysics},}\ \ (\bibinfo  {publisher} {Pergamon
  press},\ \bibinfo {address} {Oxford},\ \bibinfo {year} {1980})\BibitemShut
  {NoStop}%
\bibitem [{\citenamefont {Speth}\ \emph {et~al.}(1977)\citenamefont {Speth},
  \citenamefont {Werner},\ and\ \citenamefont {Wild}}]{Speth_1977}%
  \BibitemOpen
  \bibfield  {author} {\bibinfo {author} {\bibfnamefont {J.}~\bibnamefont
  {Speth}}, \bibinfo {author} {\bibfnamefont {E.}~\bibnamefont {Werner}}, \
  and\ \bibinfo {author} {\bibfnamefont {W.}~\bibnamefont {Wild}},\ }\href@noop
  {} {\bibfield  {journal} {\bibinfo  {journal} {Phys. Rep.}\ }\textbf
  {\bibinfo {volume} {33}},\ \bibinfo {pages} {127} (\bibinfo {year}
  {1977})}\BibitemShut {NoStop}%
\bibitem [{\citenamefont {Skyrme}(1959)}]{Sky59a}%
  \BibitemOpen
  \bibfield  {author} {\bibinfo {author} {\bibfnamefont {T.~H.~R.}\
  \bibnamefont {Skyrme}},\ }\href@noop {} {\bibfield  {journal} {\bibinfo
  {journal} {Nucl. Phys.}\ }\textbf {\bibinfo {volume} {9}},\ \bibinfo {pages}
  {615} (\bibinfo {year} {1959})}\BibitemShut {NoStop}%
\bibitem [{\citenamefont {Negele}\ and\ \citenamefont
  {Vautherin}(1972)}]{Neg72a}%
  \BibitemOpen
  \bibfield  {author} {\bibinfo {author} {\bibfnamefont {J.~W.}\ \bibnamefont
  {Negele}}\ and\ \bibinfo {author} {\bibfnamefont {D.}~\bibnamefont
  {Vautherin}},\ }\href@noop {} {\bibfield  {journal} {\bibinfo  {journal}
  {Phys. Rev. C}\ }\textbf {\bibinfo {volume} {C5}},\ \bibinfo {pages} {1472}
  (\bibinfo {year} {1972})}\BibitemShut {NoStop}%
\bibitem [{\citenamefont {Negele}\ and\ \citenamefont
  {Vautherin}(1975)}]{Neg75a}%
  \BibitemOpen
  \bibfield  {author} {\bibinfo {author} {\bibfnamefont {J.~W.}\ \bibnamefont
  {Negele}}\ and\ \bibinfo {author} {\bibfnamefont {D.}~\bibnamefont
  {Vautherin}},\ }\href@noop {} {\bibfield  {journal} {\bibinfo  {journal}
  {Phys. Rev. C}\ }\textbf {\bibinfo {volume} {C11}},\ \bibinfo {pages} {1031}
  (\bibinfo {year} {1975})}\BibitemShut {NoStop}%
\bibitem [{\citenamefont {Liu}\ and\ \citenamefont {Brown}(1976)}]{Liu76}%
  \BibitemOpen
  \bibfield  {author} {\bibinfo {author} {\bibfnamefont {K.~F.}\ \bibnamefont
  {Liu}}\ and\ \bibinfo {author} {\bibfnamefont {G.~E.}\ \bibnamefont
  {Brown}},\ }\href@noop {} {\bibfield  {journal} {\bibinfo  {journal} {Nucl.
  Phys. A}\ }\textbf {\bibinfo {volume} {265}},\ \bibinfo {pages} {385}
  (\bibinfo {year} {1976})}\BibitemShut {NoStop}%
\bibitem [{\citenamefont {Bender}\ \emph {et~al.}(2003)\citenamefont {Bender},
  \citenamefont {Heenen},\ and\ \citenamefont {Reinhard}}]{Bender_2003}%
  \BibitemOpen
  \bibfield  {author} {\bibinfo {author} {\bibfnamefont {M.}~\bibnamefont
  {Bender}}, \bibinfo {author} {\bibfnamefont {P.-H.}\ \bibnamefont {Heenen}},
  \ and\ \bibinfo {author} {\bibfnamefont {P.-G.}\ \bibnamefont {Reinhard}},\
  }\href {\doibase 10.1103/Rev.Mod.Phys.75.121} {\bibfield  {journal} {\bibinfo
   {journal} {Rev. Mod. Phys.}\ }\textbf {\bibinfo {volume} {75}},\ \bibinfo
  {pages} {121} (\bibinfo {year} {2003})}\BibitemShut {NoStop}%
\bibitem [{\citenamefont {Vautherin}\ and\ \citenamefont
  {Brink}(1972)}]{Brink72}%
  \BibitemOpen
  \bibfield  {author} {\bibinfo {author} {\bibfnamefont {D.}~\bibnamefont
  {Vautherin}}\ and\ \bibinfo {author} {\bibfnamefont {D.}~\bibnamefont
  {Brink}},\ }\href@noop {} {\bibfield  {journal} {\bibinfo  {journal} {Phys.
  Rev.C}\ }\textbf {\bibinfo {volume} {5}},\ \bibinfo {pages} {626} (\bibinfo
  {year} {1972})}\BibitemShut {NoStop}%
\bibitem [{\citenamefont {Kl{\"{u}}pfel}\ \emph {et~al.}(2009)\citenamefont
  {Kl{\"{u}}pfel}, \citenamefont {Reinhard}, \citenamefont {B{\"{u}}rvenich},\
  and\ \citenamefont {Maruhn}}]{Kluepfel_2009}%
  \BibitemOpen
  \bibfield  {author} {\bibinfo {author} {\bibfnamefont {P.}~\bibnamefont
  {Kl{\"{u}}pfel}}, \bibinfo {author} {\bibfnamefont {P.-G.}\ \bibnamefont
  {Reinhard}}, \bibinfo {author} {\bibfnamefont {T.~J.}\ \bibnamefont
  {B{\"{u}}rvenich}}, \ and\ \bibinfo {author} {\bibfnamefont {J.~A.}\
  \bibnamefont {Maruhn}},\ }\href {\doibase 10.1103/PhysRevC.79.034310}
  {\bibfield  {journal} {\bibinfo  {journal} {Phys. Rev. C}\ }\textbf {\bibinfo
  {volume} {79}},\ \bibinfo {pages} {034310} (\bibinfo {year}
  {2009})}\BibitemShut {NoStop}%
\bibitem [{\citenamefont {Bardin}\ and\ \citenamefont
  {Schrieffer}(1961)}]{Bardin_1961}%
  \BibitemOpen
  \bibfield  {author} {\bibinfo {author} {\bibfnamefont {J.}~\bibnamefont
  {Bardin}}\ and\ \bibinfo {author} {\bibfnamefont {J.~R.}\ \bibnamefont
  {Schrieffer}},\ }in\ \href@noop {} {\emph {\bibinfo {booktitle} {Progress in
  low Temperatur Physics}}},\ Vol.~\bibinfo {volume} {3},\ \bibinfo {editor}
  {edited by\ \bibinfo {editor} {\bibfnamefont {C.~J.}\ \bibnamefont
  {Gorter}}}\ (\bibinfo  {publisher} {North-Holland},\ \bibinfo {year} {1961})\
  p.\ \bibinfo {pages} {170}\BibitemShut {NoStop}%
\bibitem [{\citenamefont {Bell}\ and\ \citenamefont
  {Skyrme}(1956)}]{Bell_1956}%
  \BibitemOpen
  \bibfield  {author} {\bibinfo {author} {\bibfnamefont {J.~S.}\ \bibnamefont
  {Bell}}\ and\ \bibinfo {author} {\bibfnamefont {T.~H.~R.}\ \bibnamefont
  {Skyrme}},\ }\href@noop {} {\bibfield  {journal} {\bibinfo  {journal}
  {Philos. Mag.}\ }\textbf {\bibinfo {volume} {1}},\ \bibinfo {pages} {1055}
  (\bibinfo {year} {1956})}\BibitemShut {NoStop}%
\bibitem [{\citenamefont {Van~Giai}\ and\ \citenamefont
  {Sagawa}(1981)}]{vanGiai_1981}%
  \BibitemOpen
  \bibfield  {author} {\bibinfo {author} {\bibfnamefont {N.}~\bibnamefont
  {Van~Giai}}\ and\ \bibinfo {author} {\bibfnamefont {H.}~\bibnamefont
  {Sagawa}},\ }\href@noop {} {\bibfield  {journal} {\bibinfo  {journal} {Phys.
  Lett. B}\ }\textbf {\bibinfo {volume} {106}},\ \bibinfo {pages} {379}
  (\bibinfo {year} {1981})}\BibitemShut {NoStop}%
\bibitem [{\citenamefont {Vesely}\ \emph {et~al.}(2009)\citenamefont {Vesely},
  \citenamefont {Kvasil}, \citenamefont {Nesterenko}, \citenamefont {Kleinig},
  \citenamefont {Reinhard},\ and\ \citenamefont {Ponomarev}}]{Vesely_2009}%
  \BibitemOpen
  \bibfield  {author} {\bibinfo {author} {\bibfnamefont {P.}~\bibnamefont
  {Vesely}}, \bibinfo {author} {\bibfnamefont {J.}~\bibnamefont {Kvasil}},
  \bibinfo {author} {\bibfnamefont {V.~O.}\ \bibnamefont {Nesterenko}},
  \bibinfo {author} {\bibfnamefont {W.}~\bibnamefont {Kleinig}}, \bibinfo
  {author} {\bibfnamefont {P.-G.}\ \bibnamefont {Reinhard}}, \ and\ \bibinfo
  {author} {\bibfnamefont {V.~Y.}\ \bibnamefont {Ponomarev}},\ }\href@noop {}
  {\bibfield  {journal} {\bibinfo  {journal} {Phys. Rev. C}\ }\textbf {\bibinfo
  {volume} {80}},\ \bibinfo {pages} {031302(R)} (\bibinfo {year}
  {2009})}\BibitemShut {NoStop}%
\bibitem [{\citenamefont {Nesterenko}\ \emph {et~al.}(2010)\citenamefont
  {Nesterenko}, \citenamefont {Kvasil}, \citenamefont {Vesely}, \citenamefont
  {Kleinig}, \citenamefont {Reinhard},\ and\ \citenamefont
  {Ponomarev}}]{Nesterenko_2010}%
  \BibitemOpen
  \bibfield  {author} {\bibinfo {author} {\bibfnamefont {V.~O.}\ \bibnamefont
  {Nesterenko}}, \bibinfo {author} {\bibfnamefont {J.}~\bibnamefont {Kvasil}},
  \bibinfo {author} {\bibfnamefont {P.}~\bibnamefont {Vesely}}, \bibinfo
  {author} {\bibfnamefont {W.}~\bibnamefont {Kleinig}}, \bibinfo {author}
  {\bibfnamefont {P.-G.}\ \bibnamefont {Reinhard}}, \ and\ \bibinfo {author}
  {\bibfnamefont {V.~Y.}\ \bibnamefont {Ponomarev}},\ }\href@noop {} {\bibfield
   {journal} {\bibinfo  {journal} {J. Phys. G: Nucl. Part. Phys.}\ }\textbf
  {\bibinfo {volume} {37}},\ \bibinfo {pages} {064034} (\bibinfo {year}
  {2010})}\BibitemShut {NoStop}%
\bibitem [{\citenamefont {Tselyaev}\ \emph {et~al.}(2019)\citenamefont
  {Tselyaev}, \citenamefont {Lyutorovich}, \citenamefont {Speth}, \citenamefont
  {Reinhard},\ and\ \citenamefont {Smirnov}}]{Tselyaev_2019}%
  \BibitemOpen
  \bibfield  {author} {\bibinfo {author} {\bibfnamefont {V.}~\bibnamefont
  {Tselyaev}}, \bibinfo {author} {\bibfnamefont {N.}~\bibnamefont
  {Lyutorovich}}, \bibinfo {author} {\bibfnamefont {J.}~\bibnamefont {Speth}},
  \bibinfo {author} {\bibfnamefont {P.-G.}\ \bibnamefont {Reinhard}}, \ and\
  \bibinfo {author} {\bibfnamefont {D.}~\bibnamefont {Smirnov}},\ }\href
  {https://link.aps.org/doi/1903.01585} {\bibfield  {journal} {\bibinfo
  {journal} {Phys. Rev. C}\ }\textbf {\bibinfo {volume} {99}},\ \bibinfo
  {pages} {064329} (\bibinfo {year} {2019})}\BibitemShut {NoStop}%
\bibitem [{\citenamefont {Erler}\ \emph {et~al.}(2011)\citenamefont {Erler},
  \citenamefont {Kl{\"{u}}pfel},\ and\ \citenamefont {Reinhard}}]{Erler_2011}%
  \BibitemOpen
  \bibfield  {author} {\bibinfo {author} {\bibfnamefont {J.}~\bibnamefont
  {Erler}}, \bibinfo {author} {\bibfnamefont {P.}~\bibnamefont
  {Kl{\"{u}}pfel}}, \ and\ \bibinfo {author} {\bibfnamefont {P.~G.}\
  \bibnamefont {Reinhard}},\ }\href {doi:10.1088/0954-3899/38/3/033101}
  {\bibfield  {journal} {\bibinfo  {journal} {J. Phys. G}\ }\textbf {\bibinfo
  {volume} {38}},\ \bibinfo {pages} {033101} (\bibinfo {year}
  {2011})}\BibitemShut {NoStop}%
\bibitem [{\citenamefont {Reinhard}(1992)}]{Rei92b}%
  \BibitemOpen
  \bibfield  {author} {\bibinfo {author} {\bibfnamefont {P.-G.}\ \bibnamefont
  {Reinhard}},\ }\href@noop {} {\bibfield  {journal} {\bibinfo  {journal} {Ann.
  Phys. (Leipzig)}\ }\textbf {\bibinfo {volume} {504}},\ \bibinfo {pages} {632}
  (\bibinfo {year} {1992})}\BibitemShut {NoStop}%
\bibitem [{\citenamefont {Repko}\ and\ \citenamefont {Kvasil}(2019)}]{Rep19a}%
  \BibitemOpen
  \bibfield  {author} {\bibinfo {author} {\bibfnamefont {A.}~\bibnamefont
  {Repko}}\ and\ \bibinfo {author} {\bibfnamefont {J.}~\bibnamefont {Kvasil}},\
  }\href@noop {} {\bibfield  {journal} {\bibinfo  {journal} {Acta Phys. Pol. B
  Proc. Suppl.}\ }\textbf {\bibinfo {volume} {12}},\ \bibinfo {pages} {689}
  (\bibinfo {year} {2019})}\BibitemShut {NoStop}%
\bibitem [{\citenamefont {Ring}\ and\ \citenamefont {Schuck}(1980)}]{Ring80}%
  \BibitemOpen
  \bibfield  {author} {\bibinfo {author} {\bibfnamefont {P.}~\bibnamefont
  {Ring}}\ and\ \bibinfo {author} {\bibfnamefont {P.}~\bibnamefont {Schuck}},\
  }\href@noop {} {\emph {\bibinfo {title} {The Nuclear Many-Body Problem}}}\
  (\bibinfo  {publisher} {Springer},\ \bibinfo {address} {New York, Heidelberg,
  Berlin},\ \bibinfo {year} {1980})\BibitemShut {NoStop}%
\bibitem [{\citenamefont {Pines}\ and\ \citenamefont
  {Nozi\`eres}(1966)}]{Pin66}%
  \BibitemOpen
  \bibfield  {author} {\bibinfo {author} {\bibfnamefont {D.}~\bibnamefont
  {Pines}}\ and\ \bibinfo {author} {\bibfnamefont {P.}~\bibnamefont
  {Nozi\`eres}},\ }\href@noop {} {\emph {\bibinfo {title} {{The Theory of
  Quantum Liquids}}}}\ (\bibinfo  {publisher} {W A Benjamin},\ \bibinfo
  {address} {New York},\ \bibinfo {year} {1966})\BibitemShut {NoStop}%
\bibitem [{\citenamefont {McNeil}\ \emph {et~al.}(1986)\citenamefont {McNeil},
  \citenamefont {Amado}, \citenamefont {Horowitz}, \citenamefont {Oka},
  \citenamefont {Shepard},\ and\ \citenamefont {Sparrow}}]{McN86a}%
  \BibitemOpen
  \bibfield  {author} {\bibinfo {author} {\bibfnamefont {J.~A.}\ \bibnamefont
  {McNeil}}, \bibinfo {author} {\bibfnamefont {R.~D.}\ \bibnamefont {Amado}},
  \bibinfo {author} {\bibfnamefont {C.~J.}\ \bibnamefont {Horowitz}}, \bibinfo
  {author} {\bibfnamefont {M.}~\bibnamefont {Oka}}, \bibinfo {author}
  {\bibfnamefont {J.~R.}\ \bibnamefont {Shepard}}, \ and\ \bibinfo {author}
  {\bibfnamefont {D.~A.}\ \bibnamefont {Sparrow}},\ }\href {\doibase
  10.1103/PhysRevC.34.746} {\bibfield  {journal} {\bibinfo  {journal} {Phys.
  Rev. C}\ }\textbf {\bibinfo {volume} {34}},\ \bibinfo {pages} {746} (\bibinfo
  {year} {1986})}\BibitemShut {NoStop}%
\bibitem [{\citenamefont {Landau}(1957{\natexlab{a}})}]{Landau1}%
  \BibitemOpen
  \bibfield  {author} {\bibinfo {author} {\bibfnamefont {L.~D.}\ \bibnamefont
  {Landau}},\ }\href@noop {} {\bibfield  {journal} {\bibinfo  {journal} {JETP}\
  }\textbf {\bibinfo {volume} {3}},\ \bibinfo {pages} {920} (\bibinfo {year}
  {1957}{\natexlab{a}})}\BibitemShut {NoStop}%
\bibitem [{\citenamefont {Landau}(1957{\natexlab{b}})}]{Landau2}%
  \BibitemOpen
  \bibfield  {author} {\bibinfo {author} {\bibfnamefont {L.~D.}\ \bibnamefont
  {Landau}},\ }\href@noop {} {\bibfield  {journal} {\bibinfo  {journal} {JETP}\
  }\textbf {\bibinfo {volume} {5}},\ \bibinfo {pages} {101} (\bibinfo {year}
  {1957}{\natexlab{b}})}\BibitemShut {NoStop}%
\bibitem [{\citenamefont {Landau}(1959)}]{Landau3}%
  \BibitemOpen
  \bibfield  {author} {\bibinfo {author} {\bibfnamefont {L.~D.}\ \bibnamefont
  {Landau}},\ }\href@noop {} {\bibfield  {journal} {\bibinfo  {journal} {JETP}\
  }\textbf {\bibinfo {volume} {8}},\ \bibinfo {pages} {70} (\bibinfo {year}
  {1959})}\BibitemShut {NoStop}%
\bibitem [{\citenamefont {Speth}\ \emph {et~al.}(2014)\citenamefont {Speth},
  \citenamefont {Krewald}, \citenamefont {Gr\"ummer}, \citenamefont {Reinhard},
  \citenamefont {Lyutorovich},\ and\ \citenamefont {Tselyaev}}]{Speth_2014}%
  \BibitemOpen
  \bibfield  {author} {\bibinfo {author} {\bibfnamefont {J.}~\bibnamefont
  {Speth}}, \bibinfo {author} {\bibfnamefont {S.}~\bibnamefont {Krewald}},
  \bibinfo {author} {\bibfnamefont {F.}~\bibnamefont {Gr\"ummer}}, \bibinfo
  {author} {\bibfnamefont {P.~G.}\ \bibnamefont {Reinhard}}, \bibinfo {author}
  {\bibfnamefont {N.}~\bibnamefont {Lyutorovich}}, \ and\ \bibinfo {author}
  {\bibfnamefont {V.}~\bibnamefont {Tselyaev}},\ }\href@noop {} {\bibfield
  {journal} {\bibinfo  {journal} {Nucl. Phys.}\ }\textbf {\bibinfo {volume}
  {A928}},\ \bibinfo {pages} {17} (\bibinfo {year} {2014})}\BibitemShut
  {NoStop}%
\bibitem [{\citenamefont {Lyutorovich}\ \emph {et~al.}(2012)\citenamefont
  {Lyutorovich}, \citenamefont {Tselyaev}, \citenamefont {Speth}, \citenamefont
  {Krewald}, \citenamefont {Gr{\"{u}}mmer},\ and\ \citenamefont
  {Reinhard}}]{Lyutorovich_2012}%
  \BibitemOpen
  \bibfield  {author} {\bibinfo {author} {\bibfnamefont {N.}~\bibnamefont
  {Lyutorovich}}, \bibinfo {author} {\bibfnamefont {V.~I.}\ \bibnamefont
  {Tselyaev}}, \bibinfo {author} {\bibfnamefont {J.}~\bibnamefont {Speth}},
  \bibinfo {author} {\bibfnamefont {S.}~\bibnamefont {Krewald}}, \bibinfo
  {author} {\bibfnamefont {F.}~\bibnamefont {Gr{\"{u}}mmer}}, \ and\ \bibinfo
  {author} {\bibfnamefont {P.-G.}\ \bibnamefont {Reinhard}},\ }\href {\doibase
  10.1103/PhysRevLett.109.092502} {\bibfield  {journal} {\bibinfo  {journal}
  {Phys. Rev. Lett.}\ }\textbf {\bibinfo {volume} {109}},\ \bibinfo {pages}
  {092502} (\bibinfo {year} {2012})}\BibitemShut {NoStop}%
\bibitem [{\citenamefont {Reinhard}(1989)}]{Rei89aR}%
  \BibitemOpen
  \bibfield  {author} {\bibinfo {author} {\bibfnamefont {P.-G.}\ \bibnamefont
  {Reinhard}},\ }\href {http://iopscience.iop.org/0034-4885/52/4/002}
  {\bibfield  {journal} {\bibinfo  {journal} {Rep. Prog. Phys.}\ }\textbf
  {\bibinfo {volume} {52}},\ \bibinfo {pages} {439} (\bibinfo {year}
  {1989})}\BibitemShut {NoStop}%
\bibitem [{\citenamefont {Pototzky}\ \emph {et~al.}(2010)\citenamefont
  {Pototzky}, \citenamefont {Erler}, \citenamefont {Reinhard},\ and\
  \citenamefont {Nesterenko}}]{Pot10a}%
  \BibitemOpen
  \bibfield  {author} {\bibinfo {author} {\bibfnamefont {K.~J.}\ \bibnamefont
  {Pototzky}}, \bibinfo {author} {\bibfnamefont {J.}~\bibnamefont {Erler}},
  \bibinfo {author} {\bibfnamefont {P.-G.}\ \bibnamefont {Reinhard}}, \ and\
  \bibinfo {author} {\bibfnamefont {V.~O.}\ \bibnamefont {Nesterenko}},\ }\href
  {http://dx.doi.org/10.1140/epja/i2010-11045-6} {\bibfield  {journal}
  {\bibinfo  {journal} {Eur. Phys. J. A}\ }\textbf {\bibinfo {volume} {46}},\
  \bibinfo {pages} {299} (\bibinfo {year} {2010})}\BibitemShut {NoStop}%
\bibitem [{\citenamefont {Speth}\ and\ \citenamefont
  {Wambach}(1991)}]{Speth_1991}%
  \BibitemOpen
  \bibfield  {author} {\bibinfo {author} {\bibfnamefont {J.}~\bibnamefont
  {Speth}}\ and\ \bibinfo {author} {\bibfnamefont {J.}~\bibnamefont
  {Wambach}},\ }in\ \href@noop {} {\emph {\bibinfo {booktitle} {Electric and
  Magnetic Giant Resonances in Nuclei}}},\ Vol.\ \bibinfo {volume}
  {International Review of Nuclear Physics, Vol. 7},\ \bibinfo {editor} {edited
  by\ \bibinfo {editor} {\bibfnamefont {J.}~\bibnamefont {Speth}}}\ (\bibinfo
  {publisher} {World Scientific},\ \bibinfo {year} {1991})\ pp.\ \bibinfo
  {pages} {2--87}\BibitemShut {NoStop}%
\bibitem [{\citenamefont {Tselyaev}(2007)}]{Tselyaev_2007}%
  \BibitemOpen
  \bibfield  {author} {\bibinfo {author} {\bibfnamefont {V.~I.}\ \bibnamefont
  {Tselyaev}},\ }\href {\doibase 10.1103/Phys.RevC.75.024306} {\bibfield
  {journal} {\bibinfo  {journal} {Phys. Rev. C}\ }\textbf {\bibinfo {volume}
  {75}},\ \bibinfo {pages} {024306} (\bibinfo {year} {2007})}\BibitemShut
  {NoStop}%
\bibitem [{\citenamefont {Tselyaev}\ \emph {et~al.}(2016)\citenamefont
  {Tselyaev}, \citenamefont {Lyutorovich}, \citenamefont {Speth}, \citenamefont
  {Krewald},\ and\ \citenamefont {Reinhard}}]{Tselyaev_2016}%
  \BibitemOpen
  \bibfield  {author} {\bibinfo {author} {\bibfnamefont {V.}~\bibnamefont
  {Tselyaev}}, \bibinfo {author} {\bibfnamefont {N.}~\bibnamefont
  {Lyutorovich}}, \bibinfo {author} {\bibfnamefont {J.}~\bibnamefont {Speth}},
  \bibinfo {author} {\bibfnamefont {S.}~\bibnamefont {Krewald}}, \ and\
  \bibinfo {author} {\bibfnamefont {P.-G.}\ \bibnamefont {Reinhard}},\ }\href
  {\doibase 10.1103/PhysRevC.94.034306} {\bibfield  {journal} {\bibinfo
  {journal} {Phys. Rev. C}\ }\textbf {\bibinfo {volume} {94}},\ \bibinfo
  {pages} {034306} (\bibinfo {year} {2016})}\BibitemShut {NoStop}%
\bibitem [{\citenamefont {Tselyaev}\ \emph {et~al.}(2017)\citenamefont
  {Tselyaev}, \citenamefont {Lyutorovich}, \citenamefont {Speth},\ and\
  \citenamefont {Reinhard}}]{Tselyaev_2017}%
  \BibitemOpen
  \bibfield  {author} {\bibinfo {author} {\bibfnamefont {V.}~\bibnamefont
  {Tselyaev}}, \bibinfo {author} {\bibfnamefont {N.}~\bibnamefont
  {Lyutorovich}}, \bibinfo {author} {\bibfnamefont {J.}~\bibnamefont {Speth}},
  \ and\ \bibinfo {author} {\bibfnamefont {P.-G.}\ \bibnamefont {Reinhard}},\
  }\href {\doibase 10.1103/PhysRevC.96.024312} {\bibfield  {journal} {\bibinfo
  {journal} {Phys. Rev. C}\ }\textbf {\bibinfo {volume} {96}},\ \bibinfo
  {pages} {024312} (\bibinfo {year} {2017})}\BibitemShut {NoStop}%
\bibitem [{\citenamefont {Tselyaev}\ \emph {et~al.}(2018)\citenamefont
  {Tselyaev}, \citenamefont {Lyutorovich}, \citenamefont {Speth},\ and\
  \citenamefont {Reinhard}}]{Tselyaev_2018}%
  \BibitemOpen
  \bibfield  {author} {\bibinfo {author} {\bibfnamefont {V.}~\bibnamefont
  {Tselyaev}}, \bibinfo {author} {\bibfnamefont {N.}~\bibnamefont
  {Lyutorovich}}, \bibinfo {author} {\bibfnamefont {J.}~\bibnamefont {Speth}},
  \ and\ \bibinfo {author} {\bibfnamefont {P.-G.}\ \bibnamefont {Reinhard}},\
  }\href {\doibase 10.1103/PhysRevC.97.044308} {\bibfield  {journal} {\bibinfo
  {journal} {Phys. Rev. C}\ }\textbf {\bibinfo {volume} {97}},\ \bibinfo
  {pages} {044308} (\bibinfo {year} {2018})}\BibitemShut {NoStop}%
\bibitem [{\citenamefont {Wakasa}\ \emph {et~al.}(2012)\citenamefont {Wakasa},
  \citenamefont {Okamoto}, \citenamefont {Dozono}, \citenamefont {Hatanaka},
  \citenamefont {Ichimura}, \citenamefont {Kuroita}, \citenamefont {Maeda},
  \citenamefont {Miyasako}, \citenamefont {Noro}, \citenamefont {Saito},
  \citenamefont {Sakemi}, \citenamefont {Yabe},\ and\ \citenamefont
  {Yako}}]{Wakasa_2012}%
  \BibitemOpen
  \bibfield  {author} {\bibinfo {author} {\bibfnamefont {T.}~\bibnamefont
  {Wakasa}}, \bibinfo {author} {\bibfnamefont {M.}~\bibnamefont {Okamoto}},
  \bibinfo {author} {\bibfnamefont {M.}~\bibnamefont {Dozono}}, \bibinfo
  {author} {\bibfnamefont {K.}~\bibnamefont {Hatanaka}}, \bibinfo {author}
  {\bibfnamefont {M.}~\bibnamefont {Ichimura}}, \bibinfo {author}
  {\bibfnamefont {S.}~\bibnamefont {Kuroita}}, \bibinfo {author} {\bibfnamefont
  {Y.}~\bibnamefont {Maeda}}, \bibinfo {author} {\bibfnamefont
  {H.}~\bibnamefont {Miyasako}}, \bibinfo {author} {\bibfnamefont
  {T.}~\bibnamefont {Noro}}, \bibinfo {author} {\bibfnamefont {T.}~\bibnamefont
  {Saito}}, \bibinfo {author} {\bibfnamefont {Y.}~\bibnamefont {Sakemi}},
  \bibinfo {author} {\bibfnamefont {T.}~\bibnamefont {Yabe}}, \ and\ \bibinfo
  {author} {\bibfnamefont {K.}~\bibnamefont {Yako}},\ }\href {\doibase
  10.1103/PhysRevC.85.064606} {\bibfield  {journal} {\bibinfo  {journal} {Phys.
  Rev. C}\ }\textbf {\bibinfo {volume} {85}},\ \bibinfo {pages} {064606}
  (\bibinfo {year} {2012})}\BibitemShut {NoStop}%
\bibitem [{\citenamefont {Ring}\ and\ \citenamefont
  {Speth}(1973)}]{Ring_1973plb}%
  \BibitemOpen
  \bibfield  {author} {\bibinfo {author} {\bibfnamefont {P.}~\bibnamefont
  {Ring}}\ and\ \bibinfo {author} {\bibfnamefont {J.}~\bibnamefont {Speth}},\
  }\href@noop {} {\bibfield  {journal} {\bibinfo  {journal} {Phys. Lett. B}\
  }\textbf {\bibinfo {volume} {44}},\ \bibinfo {pages} {477} (\bibinfo {year}
  {1973})}\BibitemShut {NoStop}%
\bibitem [{\citenamefont {Chabanat}\ \emph {et~al.}(1997)\citenamefont
  {Chabanat}, \citenamefont {Bonche}, \citenamefont {Haensel}, \citenamefont
  {Meyer},\ and\ \citenamefont {Schaeffer}}]{Chabanat_1997}%
  \BibitemOpen
  \bibfield  {author} {\bibinfo {author} {\bibfnamefont {E.}~\bibnamefont
  {Chabanat}}, \bibinfo {author} {\bibfnamefont {P.}~\bibnamefont {Bonche}},
  \bibinfo {author} {\bibfnamefont {P.}~\bibnamefont {Haensel}}, \bibinfo
  {author} {\bibfnamefont {J.}~\bibnamefont {Meyer}}, \ and\ \bibinfo {author}
  {\bibfnamefont {R.}~\bibnamefont {Schaeffer}},\ }\href@noop {} {\bibfield
  {journal} {\bibinfo  {journal} {Nucl. Phys. A}\ }\textbf {\bibinfo {volume}
  {627}},\ \bibinfo {pages} {710} (\bibinfo {year} {1997})}\BibitemShut
  {NoStop}%
\bibitem [{\citenamefont {Margueron}\ \emph {et~al.}(2002)\citenamefont
  {Margueron}, \citenamefont {Navarro},\ and\ \citenamefont
  {Van~Giai}}]{Margueron_2002}%
  \BibitemOpen
  \bibfield  {author} {\bibinfo {author} {\bibfnamefont {J.}~\bibnamefont
  {Margueron}}, \bibinfo {author} {\bibfnamefont {J.}~\bibnamefont {Navarro}},
  \ and\ \bibinfo {author} {\bibfnamefont {N.}~\bibnamefont {Van~Giai}},\
  }\href {\doibase 10.1103/PhysRevC.66.014303} {\bibfield  {journal} {\bibinfo
  {journal} {Phys. Rev. C}\ }\textbf {\bibinfo {volume} {66}},\ \bibinfo
  {pages} {014303} (\bibinfo {year} {2002})}\BibitemShut {NoStop}%
\bibitem [{\citenamefont {Lesinski}\ \emph {et~al.}(2007)\citenamefont
  {Lesinski}, \citenamefont {Bender}, \citenamefont {Bennaceur}, \citenamefont
  {Duguet},\ and\ \citenamefont {Meyer}}]{Lesinski_2007}%
  \BibitemOpen
  \bibfield  {author} {\bibinfo {author} {\bibfnamefont {T.}~\bibnamefont
  {Lesinski}}, \bibinfo {author} {\bibfnamefont {M.}~\bibnamefont {Bender}},
  \bibinfo {author} {\bibfnamefont {K.}~\bibnamefont {Bennaceur}}, \bibinfo
  {author} {\bibfnamefont {T.}~\bibnamefont {Duguet}}, \ and\ \bibinfo {author}
  {\bibfnamefont {J.}~\bibnamefont {Meyer}},\ }\href@noop {} {\bibfield
  {journal} {\bibinfo  {journal} {Phys. Rev. C}\ }\textbf {\bibinfo {volume}
  {76}},\ \bibinfo {pages} {014312} (\bibinfo {year} {2007})}\BibitemShut
  {NoStop}%
\end{thebibliography}%

\end{document}